\documentclass{article}

\usepackage[numbers]{natbib}
\usepackage{authblk}
\usepackage[a4paper, total={161mm,247mm}]{geometry}
\usepackage[section]{placeins} 

\usepackage[utf8]{inputenc} 
\usepackage[T1]{fontenc}    
\usepackage{hyperref}       
\usepackage{url}            
\usepackage{booktabs}       
\usepackage{amsfonts}       
\usepackage{nicefrac}       
\usepackage{microtype}      
\usepackage{xcolor}         
\usepackage{amsmath}
\usepackage{amsthm} 
\usepackage{bm} 
\usepackage{mathabx} 
\usepackage{caption}
\usepackage{graphicx}
\usepackage[all,cmtip]{xy}
\usepackage{multirow} 

\usepackage[shortlabels]{enumitem} 

\DeclareMathOperator{\EX}{\mathbb{E}}

\newcommand{\indep}{\perp \!\!\! \perp}

\usepackage{thmtools,thm-restate}
\newtheorem{theorem}{Theorem}
\newtheorem{theoremi}{Theorem (informal)}
\newtheorem{lemma}{Lemma}

\usepackage[acronym,nowarn]{glossaries}
\makeglossaries
\newacronym{rct}{RCT}{randomized controlled trial}
\newacronym{cate}{CATE}{conditional average treatment effect}
\newacronym{crate}{CR-ATE}{constant-relative ATE}
\newacronym{ate}{ATE}{average treatment effect}
\newacronym{ateb}{ATE-baseline}{average treatment effect-baseline}
\newacronym{pehe}{PEHE}{precision in heterogeneous treatment effect estimation}
\newacronym{mcm}{MCM}{marginally constrained model}
\newacronym{crmcm}{CR-MCM}{constant-relative marginally constrained model}
\glsdisablehyper

\title{Conditional average treatment effect estimation with marginally constrained models}

\author[1,2]{Wouter A.C. van Amsterdam}
\author[3]{Rajesh Ranganath}
\affil[1]{
	Department of Data Science and Biostatistics\\
	Julius Center for Health Sciences and Primary Care\\
	University Medical Center Utrecht
}
\affil[2]{
	Babylon Health Inc.
}
\affil[3]{
	Courant Institute of Mathematical Sciences, Center for Data Science\\
	New York University
}

\begin{document}

\maketitle
 
  \begin{abstract}
	  {Treatment effect estimates are often available from randomized controlled trials as a single \textit{average treatment effect} for a certain patient population. Estimates of the \textit{conditional average treatment effect } (CATE) are more useful for individualized treatment decision making, but randomized trials are often too small to estimate the CATE. 
	  Examples in medical literature make use of the \textit{relative} treatment effect (e.g. an odds-ratio) reported by randomized trials to estimate the CATE using large observational datasets. 
	  One approach to estimating these CATE models is by using the relative treatment effect as an \textit{offset}, while estimating the covariate-specific untreated risk. 
	  We observe that the odds-ratios reported in randomized controlled trials are not the odds-ratios that are needed in offset models because trials often report the \textit{marginal} odds-ratio. 
	  We introduce a constraint or regularizer to better use marginal odds-ratios from randomized controlled trials and find that under the standard observational causal inference assumptions this approach provides a consistent estimate of the CATE. 
	  Next, we show that the offset approach is not valid for CATE estimation in the presence of unobserved confounding.
	  We study if the offset assumption and the marginal constraint lead to better approximations of the CATE relative to the alternative of using the average treatment effect estimate from the randomized trial.
	  We empirically show that when the underlying CATE has sufficient variation, the constraint and offset approaches lead to closer approximations to the CATE.
  }
  \end{abstract}

\section{Introduction}
Weighing potential benefits and harms of treatment requires knowing the treatment effect, which is the change in probability of an outcome between different treatments.
The gold standard for estimating treatment effects are \glspl{rct}.
\Glspl{rct} are designed to provide estimates of the \gls{ate}, which reveals if a treatment has an effect on an outcome. The \gls{ate} does not reveal \emph{which patients} would benefit from a treatment.
Tailoring the effect to a patient requires knowing how 
treatment effects change given characteristics of that patient; this is the \gls{cate}. The \gls{cate} is a measure of absolute risk, which patients prefer when making decisions \citep{murray_patients_2018}. 
Estimating \glspl{cate} directly from \gls{rct} data is often infeasible because trials are generally only powered to estimate population average effects.

Under the standard assumptions like ignorability and positivity, observational data provides an avenue for estimating \glspl{cate} \citep{pearl_causal_2009}.
However, techniques for observational causal inference often do not make use of knowing the effect provided by an \gls{rct}, which are typically reported on a relative scale as an odds-ratio or hazard-ratio (e.g. \citep{furie_two-year_2020,lean_primary_2018}).
To make use of population-level relative effects reported by \glspl{rct} in \gls{cate} estimation, 
several previous studies on breast cancer and cardiovascular disease used the assumption of a \emph{constant relative treatment effect} to develop \gls{cate} models from observational data \citep{candido_dos_reis_updated_2017, ravdin_computer_2001, alaa_machine_2021, xu_prediction_2021}.
We call this assumption the \gls{crate} assumption and models that use this assumption \gls{crate} models.

The assumption of a constant relative treatment effect does not imply the \gls{cate} must be constant
because even with a constant relative treatment effect, the treatment can have a varying effect on an absolute risk scale depending on the untreated risk of a patient.
For instance, assume that a new cholesterol lowering drug reduces the risk of cardiovascular death within the next 10 years with an odds-ratio of 0.5.
A 60-year-old male smoker with hypertension and raised cholesterol has an untreated risk of cardiovascular death of 40\% and should expect a reduction in risk of 15\% points.
A 50-year-old female without hypertension has an untreated risk of under 1\% and will have a less than 0.5\% points reduction in risk.
Given these widely different effects on an absolute probability scale, one may recommend the new cholesterol lowering drug to the 60-year-old male but not the 50-year-old female.

One approach to use the \gls{crate} assumption with observational data is to 
use the constant relative treatment effect as an \textit{offset} term in a model. 
Some models built with an offset term were found to be accurate in observational validation studies, on the basis of which treatment guidelines acknowledged a place for them in clinical decision making \citep{cardoso_early_2019, gradishar_nccn_2021}.
Because \gls{crate} models target interventional distributions, the use of such model implicitly assumes that the \gls{rct}'s relative effect provides the correct offset and that its use controls for unobserved confounding. 
However, whether the assumption is correct has not been discussed or verified.

In this work we evaluate the validity of the assumption that a known constant odds-ratio for treatment allows for \gls{cate} estimation when using the known odds-ratio as an offset term.
Under the standard requirements for observational causal inference, we show there is a mismatch in that the odds-ratios reported in \glspl{rct} are not the odds-ratios that are needed for the offset method because \glspl{rct} generally report estimates of the \textit{marginal} odds-ratio, whereas the offset method requires the \textit{conditional} odds-ratio.
Further, when there is unobserved confounding, we demonstrate that even with the correct conditional odds-ratio offsets are not sufficient for estimating \glspl{cate}.

To address the mismatch in odds-ratios, we introduce a new estimator called the \gls{mcm}. The \gls{mcm} constrains a model's implied marginal odds-ratio to be close to that reported from a \gls{rct}. We show that with no unobserved confounding this approach is a consistent estimator of the \gls{cate}, and we empirically show a large reduction in the number of samples needed to estimate the \gls{cate} when using this marginal odds-ratio constraint. We also show how this regularizer can be combined with the \gls{crate} assumption in \glsreset{crmcm}\gls{crmcm} models.

Next we turn to violations of ignorability, i.e. the assumption of no unobserved confounding. Without ignorability in offset models, consistent estimation of \glspl{cate} is not possible. However, the goal here is to study if models like the \gls{crmcm} approximate the \gls{cate} better than the alternative provided by randomized trials, the \gls{ate}. The measure of better 
is the \gls{pehe} \cite{hill_bayesian_2011}. For example, the \gls{ate} is a good approximation of the \gls{cate} when the underlying \gls{cate} has little variation as a function of the conditioning set.
Better approximations to the \gls{cate} can lead to patient benefit as the \gls{cate} leads to better individualized treatment decisions.
We empirically show in the absence of ignorability that \glspl{mcm} almost always lead to better \gls{cate} estimation than the \gls{ateb} whenever there is sufficient variation in the underlying \gls{cate}.
Finally, we demonstrate the utility of the \gls{crate} assumption and show that when the assumption holds, \glspl{crmcm} are even better than \glspl{mcm}.

\section{Offset models: estimation and non-collapsibility}

We consider models for the absolute difference in probability of a binary outcome $Y$ under two possible treatments $T \in \{0,1\}$ conditional on a pre-treatment covariate vector $X$, using data from the observational distribution $q(Y,T,X)$.
The estimation target is the \gls{cate}, conditional on $X$.
Without loss of generality, treatment $T=0$ is assumed to be the baseline treatment (or no treatment depending on the clinical context) and $T=1$ is the comparator treatment of interest.
We denote $Y_t$ as the potential outcome of $Y$ if $T$ is set to $t$ by intervention,
and $Y_t|x$ as the analogous potential outcome conditional on $X=x$.
We refer to $\Pr(Y_0=1|X)$ as the \emph{untreated risk}, meaning the probability of experiencing the outcome when assigned no treatment (or the control treatment), conditional on $X$.
The \gls{cate} is defined as:

\begin{equation}
	\label{eq:cate}
	\text{CATE} (x) := \Pr(Y_1=1 | x) - \Pr(Y_0=1|x)
\end{equation}

Taking the expected value over $X$ of the \gls{cate}, we get the \gls{ate}. Working backwards from the definition of the \gls{ate}:

\begin{align}
	\text{ATE} &= \EX[ Y_1 - Y_0] \nonumber \\
			&= \Pr(Y_1=1)  -  \Pr(Y_0=1) \nonumber  \\
		   &= \EX [\Pr(Y_1=1|x)] - \EX[ \Pr(Y_0=1|x)] \nonumber  \\
		   &= \EX \left[ \Pr(Y_1=1|x) -  \Pr(Y_0=1|x) \right] \nonumber  \\
		   &= \EX \left[\text{CATE}(x)\right]
		   \label{eq:ate}
\end{align}

The \gls{ate} (Equation \ref{eq:ate}) is \emph{one} measure of treatment effect based on the \emph{marginal} potential outcomes $Y_0,Y_1$.
In addition to the \gls{ate} expressed in absolute probabilities, a common measure of treatment effect for binary outcomes employed by \glspl{rct} is the \emph{marginal odds-ratio} (e.g. \citep{furie_two-year_2020, lean_primary_2018, suverein_early_2023,hill_methylprednisolone_2022,weghuber_once-weekly_2022}):

\begin{equation}
	\label{eq:or}
	\text{OR}(Y_1,Y_0) = \frac{\Pr(Y_1=1) (1 - \Pr(Y_0 = 1))}{(1-\Pr(Y_1=1)) \Pr(Y_0=1)}
\end{equation}

It is often more convenient to work with the log odds-ratio.
Writing $\sigma(x) = (1+e^{-x})^{-1}$ as the \emph{sigmoid} function (also known as the \emph{logistic} function) with its inverse $\sigma^{-1}$ (known as the \emph{logit} function), the log odds-ratio denoted as $\gamma$ is:

\begin{equation}
	\label{eq:lor}
	\gamma = \gamma(Y_1,Y_0) = \log \text{OR}(Y_1,Y_0) = \sigma^{-1} (\Pr(Y_1=1)) - \sigma^{-1} (\Pr(Y_0=1))
\end{equation}

The log odds-ratio $\gamma$ as a function of potential outcomes $Y_0,Y_1$ is a measure of causal treatment effect.
Throughout we will assume to have access to $\gamma$ from a \gls{rct} conducted in the same population as the observational study.
We assume that this \gls{rct} provides an unbiased estimate of the treatment effect measure $\gamma$ with infinite precision.
However, as is typically the case due to data-sharing restrictions, only $\gamma$ is available from the \gls{rct} and not the original data.
Though \glspl{rct} can estimate many other parameters of $Y_t|X$ such as the conditional odds-ratio \citep{tchetgen_tchetgen_doubly_2010,tchetgen_tchetgen_double-robust_2011}, in practice they often only report a marginal effect.
In the discussion section we describe ways to relax these assumptions.

\subsection{Offset models as CATE models}

We now study of a class of \glsreset{crate}\gls{crate} models that are already used in clinical practice: \emph{offset models} \citep{candido_dos_reis_updated_2017}.
Relying on an assumed functional form of $\Pr(Y_t|X)$, offset models aim to approximate $\Pr(Y_t|X)$ in a constrained way by forcing the `effect' of treatment to be equal to some known, constant, \emph{relative treatment effect} on an appropriate scale.
In the context of binary outcomes, one \gls{crate} assumption is that the odds-ratio for treatment is constant for all $X$, whereas the \emph{untreated} risk $\Pr(Y_0=1|X)$ varies with $X$, meaning that:
\begin{equation}
	\label{eq:offset_assumption}
	\Pr(Y_t=1|x) = \sigma \left( \sigma^{-1}(\Pr(Y_0=1|x)) + \beta_t t\right).
\end{equation}
The assumption in Equation \ref{eq:offset_assumption} is called the \emph{offset assumption} for logistic models and is an example of a constant-relative treatment effect assumption.
Note that $\beta_t$ is a measure of treatment effect derived from the conditional potential outcomes $Y_0|X,Y_1|X$, and the specific offset assumption is that $\beta_t$ does not depend on $X$.
Using this assumption with also assuming $\beta_t$ is known yields a class of models of $\Pr(Y_t=1|X)$ called $g: \{0,1\} \bigtimes \mathcal{X} \to [0,1]$, defined as:

\begin{equation}
	\label{eq:offset_model}
	g(t,x) = \sigma \left( \sigma^{-1}(g_0(x)) + \beta_t t \right),
\end{equation}
where $g_0 = g(0,x): \mathcal{X} \to [0,1]$.
In the context of generalized linear models, a fixed term in a model that is not estimated from data is called an \textit{offset} term \citep{watson_practitioners_2007}. 
We therefore refer to models of the form of Equation \ref{eq:offset_model} as \textit{treatment offset models} or \textit{offset models} for short.
Later we will study under what conditions offset models lead to consistent estimators of the \gls{cate}.

Offset models and analogous \gls{crate} models have been used in practice without any justification \citep{candido_dos_reis_updated_2017, ravdin_computer_2001, alaa_machine_2021, xu_prediction_2021}.
Though the parametric assumption in Equation \ref{eq:offset_assumption} may seem strong, one supporting argument is that the odds-ratio was a more stable measure of treatment effect than the absolute risk difference in a review of 125 meta-analyses of \glspl{rct} \citep{engels_heterogeneity_2000}.
Additionally, \glspl{rct} rarely find evidence for variation in the odds-ratio depending on covariates (i.e. interaction terms between covariates and treatment on the log-odds scale), though it should be noted that \glspl{rct} are generally underpowered for estimating these interaction terms.
Either way, the fact that offset models are used in practice warrants a formal understanding of their consistency and under what conditions they are likely to provide a better approximation to the \gls{cate} than the \gls{ateb}.

\subsection{Estimation of offset models}

To study offset models we must first describe how they may be estimated.
One can construct an estimator for offset models of the form in Equation \ref{eq:offset_model} by specifying a class of functions $\mathcal{G}_0 = \left\{g_0 \in \mathcal{G}_0: \mathcal{X} \to [0,1] \right\}$ for the untreated risk $\Pr(Y_0=1|X)$.
Assume that $X$ is discrete with cardinality $d$.
Coding $X$ as a $d$-dimensional one-hot vector leads to a natural parameterization for non-parametric models of $\Pr(Y_0=1|X)$ with $d$ logistic regression parameters $\bm{b} \in \mathbb{R}^d$,
giving rise to model family $\mathcal{G}_0 = \{g_0: \mathcal{X} \to [0,1], g_0(x;\bm{b}) = \sigma(\bm{b}' x), \bm{b} \in \mathbb{R}^d\}$.
This $\mathcal{G}_0$ together with a known $\beta_t$ defines a family of offset models:

\begin{equation}
	\label{eq:offset_family}
	\mathcal{G} = \left\{ g: \{0,1\} \bigtimes \mathcal{X} \to [0,1], g(t,x;\bm{b},\beta_t) = \sigma \left( \bm{b}' x + \beta_t t \right), \bm{b} \in \mathbb{R}^d \right\}
\end{equation}

Parameter vector $\bm{b}$ may be obtained by maximizing the likelihood of the observed data.
The question is under what conditions this yields a consistent estimator of $\Pr(Y_t=1|X)$.
If for every $x \in \mathcal{X}, 0<\Pr(Y_0=1|X=x)<1$, there is always a $\bm{b}^{*} \in \mathbb{R}^d$ such that $\Pr(Y_0=1|x) = \sigma ( \bm{b}^{*\prime} x)$.
With this $\bm{b}^*$ we have that by the offset assumption:

\begin{align*}
	g^*(t,x ; \bm{b}^*) &:= \sigma \left( \bm{b}^{*\prime} x + \beta_t t \right) \\
		 &= \sigma \left( \sigma^{-1}(\Pr(Y_0=1|x)) + \beta_t t \right) \\
		 &= \Pr(Y_t=1|x)
\end{align*}

\paragraph{Collapsibility.}
An important concept when considering offset models is \emph{collapsibility}.
A measure $\mu(Y_0,Y_1)$ of causal effect is said to be \emph{collapsible} over variable $X$ if there exist a set of weights $w_x$ such that $\mu(Y_0,Y_1)=\frac{1}{\sum w_x} \sum w_x \mu(Y_0,Y_1|x)$, meaning that the marginal effect is a weighted average of the $x$-conditional effects \citep{huitfeldt_collapsibility_2019, didelez_logic_2022}.
As shown in Equation \ref{eq:ate} the \gls{ate} is a collapsible effect measure by weighting the \glspl{cate} by the probability of $X$.
For odds-ratios, except for very special circumstances, there are no such weights meaning that the odds-ratio is \emph{non-collapsible} \citep{whittemore_collapsibility_1978,daniel_making_2021}.

The non-collapsibility of the odds-ratio creates a problem for estimating logistic offset models as the \emph{marginal log odds-ratio} $\gamma$ reported in the \glspl{rct} is not equal to the \emph{conditional log odds-ratio} $\beta_t$ required in Equation \ref{eq:offset_model}.
This means that the model in Equation \ref{eq:offset_model} cannot be estimated from the available data.
The stronger the $T$-conditional association between $X$ and $Y$, the greater the difference between $\gamma$ and $\beta_t$ \citep{hauck_consequence_1991}.
For an illustration, see Appendix \ref{app:nc}.
This mismatch is an important drawback as at the same time, a stronger association between $X$ and $Y$ conditional on $T$ results in more variation in $\Pr(Y_0|X)$ and thus more variation in the \gls{cate} under the constant-relative treatment effect assumption.
So in the situation where offset models have more potential added value (when the \glspl{cate} vary substantially), the estimate of the marginal log odds-ratio $\gamma$ from \glspl{rct} becomes a less accurate approximation of the conditional odds-ratio $\beta_t$ needed for estimating the offset model.
If one were to use $\gamma$ in the place of $\beta_t$ in Equation \ref{eq:offset_model}, this would lead to an inconsistent estimator.
We provide a numerical example in the experiments section (Section \ref{sec:exp_collapsibility}) that highlights this effect of non-collapsibility on offset models.

\section{Marginally Constrained Models}
We now turn to a new class of \gls{cate} models that also exploits knowledge from prior \glspl{rct}.
There are many instances where an estimate of $\gamma$ is available from the published results of an \gls{rct}, but not the \gls{cate} because the sample size of the \gls{rct} was too small.
Running a new, bigger \gls{rct} may be infeasible due to costs, or deemed unethical because of the absence of equipoise, i.e. uncertainty about what treatment is superior if the prior \gls{rct} demonstrated that one treatment was superior over the other \emph{on average}.
When turning to observational data to estimate the \glspl{cate}, instead of ignoring the $\gamma$ estimate from the \gls{rct}, we can use it in the estimation as a constraint.
We describe a procedure for incorporating $\gamma$ in \gls{cate} estimation based on the marginalization of predicted outcome probabilities.
Under some regularity conditions, we prove that this method leads to a consistent estimator of the \gls{cate} when the standard causal inference assumptions hold.

\subsection{Exploiting RCT evidence by using the marginal odds-ratio as a constraint}

Assume we have a function $g$ from a family of functions $\mathcal{G} = \left\{ g: \{0,1\} \bigtimes \mathcal{X} \to [0,1] \right\}$ where we interpret $g(t,x)$ as a predicted probability $\Pr(Y=1|t,x)=g(t,x)$.
Given a distribution $q$ over $X$, we can calculate the marginalized predictions of $g$: $\Pr(Y=1|t) = \EX_{x \sim q(x)} g(t,x)$.
Correspondingly we have the \emph{implied marginal log odds-ratio} $M(g)$ (implied by $g$ and $q$, but suppressing the dependency on $q$ in the notation) by using the marginalized predictions of $g$ over $q$.

\begin{equation}
	\label{eq:implied_gamma}
	M(g) = \sigma^{-1}(\EX g(1,X)) - \sigma^{-1}(\EX g(0,X))
\end{equation}

Given an independent and identically distributed sample of $X$ with sample size $n$, define the empirical counterpart of $M$ as:

\begin{equation}
	\label{eq:implied_gamma_empirical}
	M_n(g) = \sigma^{-1}(\frac{1}{n} \sum_{i=1}^n g(1,x_i) ) - \sigma^{-1}(\frac{1}{n} \sum_{i=1}^n g(0,x_i))
\end{equation}

This empirical marginalizer was used in \citep{daniel_making_2021}.
Assume $g$ was found by maximizing the log-likelihood of the observed data over function class $\mathcal{G}$, $L_n: \mathcal{G} \to \mathbb{R}$.
We can augment the optimization objective using the known marginal odds-ratio $\gamma$ from the prior \gls{rct} as follows:

\begin{equation}
	\label{eq:lagrangian}
	\mathcal{L}_n(g) = L_n(g) - \lambda (M_n(g) - \gamma)^2, \lambda > 0
\end{equation}

We call optimization with the objective a \emph{\acrlong{mcm}} (\gls{mcm}).
If $\mathcal{G}$ encompasses all conditional distributions on $\{0,1\} \bigtimes \mathcal{X}$, i.e. in the non-parametric estimation setting, it follows that $\exists g* \in \mathcal{G}$ such that $g^*(t,x) = \Pr(Y_t=1|x)$.
Assume that additionally the standard observational causal inference assumptions hold, namely:
$(Y_1,Y_0) \indep T|X$ (\emph{strong ignorability}),
$0<q(T|X)<1$ (\emph{positivity})
and $Y_t = Y$ if $T=t$ (\emph{consistency}).
For more details on these assumptions see e.g. \citep{pearl_overview_2009, hernan_causal_nodate}.
Theorem \ref{th:constr} states that under these conditions, the augmented objective \ref{eq:lagrangian} yields a consistent estimator of $\Pr(Y_t=1|X)$.

\begin{theoremi}
	\label{th:constr}
	Given binary treatment $T$, binary outcome $Y$ and covariate $X$.
	Assume strong ignorability $(Y_1,Y_0) \indep T|X$, positivity $0<q(T|X)<1$ and consistency $Y_t=Y$ if $T=t$.
	Given a family of functions $\mathcal{G} = \left\{ g \in \mathcal{G}: \{0,1\} \bigtimes \mathcal{X} \to [0,1] \right\}$ and assuming $\exists g^* \in \mathcal{G}, \Pr(Y_t=1|x) = g^*(t,x)$.
	Denote the sample log-likelihood $L_n: \mathcal{G} \to \mathbb{R}$.
	In addition, given marginal log odds-ratio $\gamma = \sigma^{-1}(\Pr(Y_1=1)) - \sigma^{-1}(\Pr(Y_0=1))$, sample marginalizer $M_n: \mathcal{G} \to \mathbb{R}$ and $\lambda >0 $.
	Then:

	$g_n = \arg \max_{g \in \mathcal{G}} \left[ L_n(g) - \lambda ( M_n(g) - \gamma)^2 \right]$ is a consistent estimator of $\Pr(Y_t=1|X)$
	
\end{theoremi}

For proving consistency, additional assumptions on convergence and identifiability of the observational objective are required.
The formal theorem statement and proof are provided in the Appendix \ref{app:proofconsistency}. 
Under the conditions of Theorem~\ref{th:constr}, both the unconstrained maximum-likelihood estimator and the marginally constrained estimator are consistent estimators of the \gls{cate}.
However, as the \gls{mcm} optimizes over a smaller family of functions, we should expect it to be more statistically efficient.
We test the relative efficiency of the \gls{mcm} versus the unconstrained estimator in the experiments section \ref{sec:exp_efficiency}.

\subsection{Marginally constrained constant-relative treatment effect model estimation}
\Gls{mcm} leverages knowledge of the \emph{marginal} odds-ratio $\gamma$ for \gls{cate} estimation but it does not use the offset-assumption from Equation \ref{eq:offset_assumption}.
Offset models do rely on this assumption but often cannot be estimated because the required \emph{conditional} odds-ratio is not known.
We now introduce the \acrfull{crmcm}, a \gls{cate} model that leverages both the offset assumption and a known \emph{marginal} odds-ratio $\gamma$.
With discrete $X$ of cardinality $d$, non-parametric modeling of $\Pr(Y_t|X)$ requires $2d$ parameters, $d$ parameters for $\Pr(Y_0|X)$ and $d$ additional parameters for $\Pr(Y_1|X)$.
The offset assumption in Equation~\ref{eq:offset_assumption} implies that there exists a $\bm{b}^* \in \mathbb{R}^d$ and $\beta_t$ such that $\sigma \left( \bm{b}^{*\prime} x + \beta_t t \right) = \Pr(Y_t=1|x)$.
To exploit the offset assumption in \glspl{mcm} and reduce the number of parameters, we can use the constrained objective in Equation \ref{eq:lagrangian} and consider optimizing over the offset model family in Equation \ref{eq:offset_family}.
However, without access to $\beta_t$, the model family is underspecified and we cannot proceed with optimization.
Instead we introduce \gls{crmcm} as:

\begin{equation}
	\label{eq:crmcm}
	\bm{b},b_t = \arg \max_{\bm{b} \in \mathbb{R}^d, b_t \in \mathbb{R}} \left[ L_n \left( g(t,x; \bm{b},b_t) \right) - \lambda \left(M_n\left(g(t,x;\bm{b},b_t) \right) - \gamma \right)^2 \right], \lambda > 0
\end{equation}

The model family for \gls{crmcm} has $d+1$ parameters and by the offset assumption $\exists \bm{b}^*,b_t^* \in \mathbb{R}^d \bigtimes \mathbb{R}$, $g(t,x; \bm{b}^*,b_t^*) = \Pr(Y_t=1|x)$.
Because this model family is a correctly specified generalized linear model, maximizing the (unconstrained) likelihood over this family is a consistent estimator of $\Pr(Y=1|T,X)$.
Given strong ignorability, positivity and consistency, $\Pr(Y=1|T,X) = \Pr(Y_t=1|X)$.
Under the regularity conditions of Theorem~\ref{th:constr} we have that Equation \ref{eq:crmcm} is a also consistent estimator of $\Pr(Y|T,X)$.
Moreover, \gls{crmcm} may be more efficient than \gls{mcm} when the offset assumption holds.

\section{Offset models and marginally constrained models for CATE approximation under unobserved confounding}

We now turn to the case where the standard assumption of strong ignorability does not hold: in the presence of unobserved confounding.
In the presence of unobserved confounding, meaning that $Y_t$ is not independent of $T$ conditional on $X$, \gls{cate} estimators based on adjustment cannot be used.
However, this does not preclude causal effect estimation as there are known settings where causal effects can be determined despite the presence of unobserved confounding if one can rely on additional assumptions regarding the data generating mechanism.
Example methods are instrumental variable estimation \citep{wald_fitting_1940, hartford2017deep, puli2020general} and methods based on proxy-variables of unmeasured confounders \citep{miao_identifying_2018,van_amsterdam_individual_2022}.

\subsection{Offset models under unobserved confounding}

We investigate whether the structural assumption made in the offset model in Equation \ref{eq:offset_assumption} combined with knowledge of the conditional odds-ratio $\beta_t$ makes the offset model a consistent estimator of the \gls{cate}.
With a simple counter example, we prove that this is not the case.

\paragraph{Example 1: Offset models are inconsistent estimators in the presence of unobserved confounding}
\label{sec:example1}

A simple example compatible with the offset assumption in Equation \ref{eq:offset_assumption} is when there is a binary unobserved confounder $U$ but no covariate $X$.
Denoting $\mathcal{B}$ as the Bernoulli distribution and $q_u=\Pr(U=1)$, then the data-generating mechanism for this example is:

\begin{equation}\label{eq:binaryu}
	u \sim \mathcal{B}(q_u), t \sim \mathcal{B}(\Pr(T=1|u)), y \sim \mathcal{B}(\Pr(Y_t=1|u))
\end{equation}

Given the offset model family from Equation \ref{eq:offset_family}, a natural parameterization of $g(t,x)=g(t)$ in the context of this example is $g(t;b_0) = \sigma (b_0 + \beta_t t), b_0 \in \mathbb{R}$.
To disentangle the issue of non-collapsibility from unobserved confounding, we assume that $\beta_t$ is given a-priori and is not estimated.
We derive a closed-form expression for the expected log-likelihood of the observational data depending on the single parameter $b_0$ of the offset model $L(b_0)$ in the Appendix \ref{app:proofnotstationary}.
Taking the derivative with respect to $b_0$ and plugging in the ground truth value $\beta_0 := \sigma^{-1} (\Pr(Y_0=1))$ we find the following expression:

\begin{align*}
	\frac{\partial L}{\partial b_0}\bigl(b_0 = \beta_0\bigr) = q_u (1-q_u) \bigl[ & (\Pr(Y|T=0,U=1) - \Pr(Y|T=0,U=0)) \left( q(T=0|U=1) - q(T=0|U=0) \right) + \\
									    & (\Pr(Y|T=1,U=1) - \Pr(Y|T=1,U=0)) \left( q(T=1|U=1) - q(T=1|U=0) \right) \bigr]
\end{align*}

In general this expression is non-zero, meaning that the ground truth solution $\beta_0$ is not a stationary point of the expected log-likelihood.
Thereby the offset method is an inconsistent estimator for $\Pr(Y_0=1)$ in the presence of unobserved confounding.
When either $q(T | U=1) = q(T | U=0)$ or $\Pr(Y_t=1|U=1) = \Pr(Y_t=1|U=0)$, the derivative is zero at $\beta_0$, meaning that $\beta_0$ is a stationary point of the expected log-likelihood.
Either of these cases imply no unobserved confounding.

Despite its simplicity this example is important for all offset models with discrete $X$ as a)
when the treatment is binary, any arbitrary unobserved confounder can be modeled as a single binary variable while maintaining the same observational and interventional distributions \citep{ilse_combining_2022};
and b)
given $X$ with cardinality $d>1$, optimizing over the offset model family in Equation \ref{eq:offset_family} is equivalent to 
stratifying the population for each value of $X$ and optimizing the same objective as in Example 1 in each stratum.
Thus if the offset model is an inconsistent approximator of $\Pr(Y_0=1)$ in this simple example with no covariate $X$,
optimization over the offset model family for discrete $X$ will also be inconsistent for $\Pr(Y_0|X)$ by implication,
even when the correct conditional log odds-ratio $\beta_t$ is known.
In the Appendix \ref{app:proofnotstationary} we show that this implies that the offset model is inconsistent for the \gls{cate} as well, aside from very rare chance occasions.

\subsection{Metric and average treatment effect baseline}

Offset models are not consistent estimators of the \gls{cate} in the presence of unobserved confounding and require knowledge of the conditional odds-ratio which is generally not available.
However, offset models may still be useful for treatment decisions if they lead to better \gls{cate} approximation than what is used in current clinical practice.
As \glspl{rct} generally only estimate a single \acrfull{ate}, this is often what current treatment decisions are based on.
Ideally, treatment decisions are based on the difference in probability of the outcome under different treatments conditional on patient characteristics, i.e. the \gls{cate}.
A direct measure for how well a model approximates the \gls{cate} is the root-mean-squared error of approximated versus actual \gls{cate}.
In the context of \gls{cate} approximation this metric is sometimes called the \acrlong{pehe} (\gls{pehe}, \citep{hill_bayesian_2011}).
The \gls{pehe} for function $g: \{0,1\} \bigtimes \mathcal{X} \to [0,1]$ is defined as:

\begin{equation}
	\label{eq:pehe}
	\text{PEHE}(g) = \sqrt{ \EX \left( \text{CATE}(x) - (g(1,x) - g(0,x)) \right)^2}
\end{equation}

As \glspl{rct} only provide the \gls{ate}, in our experiments, we use the \gls{pehe} of the \gls{ate} as the baseline.
If the \gls{cate} is not constant but varies with $X$, the \gls{ate} is a bad approximator of the \gls{cate}, leading to suboptimal \gls{pehe} and thus suboptimal treatment decisions.
Measuring the approximation error with the \gls{pehe} is how we arrive at what we call the \gls{ateb}: the \gls{pehe} that is obtained in the current situation by using a single treatment effect for treatment decisions in all patients, instead of using the \gls{cate}.
The \gls{pehe} of the \gls{ateb} is calculated as:

\begin{equation}
	\label{eq:ateb}
	\text{PEHE}(\text{ATE}) = \sqrt{\EX \left[ \left( \text{CATE}(x)  - \text{ATE} \right)^2 \right]}
\end{equation}

Given that $\text{ATE} = \EX [\text{CATE}(x)]$, the \gls{ateb} has an intuitive form:

\begin{align}
	\text{PEHE}(\text{ATE}) &= \sqrt{\EX \left[ \left( \text{CATE}(x)  - \text{ATE} \right)^2 \right] } \nonumber \\
				&= \sqrt{\EX \left[ \left( \text{CATE}(x)  - \EX [ \text{CATE}(x) ] \right)^2 \right]  } \nonumber  \\
				&= \sqrt{\text{VAR(CATE)}} \nonumber \\
				&= \text{SD(CATE)}
	\label{eq:varcate}
\end{align}

For \gls{cate} approximation models such as the offset model and the \gls{mcm} and \gls{crmcm},
even if the model is inconsistent under certain conditions,
it may still be a valid modeling choice if it has lower \gls{pehe} than the \gls{ateb} because it can lead to better treatment decisions compared with current clinical practice.

\section{Experiments}

We now study the different models for \gls{cate} approximation in four experiments.
In the Appendix (\ref{app:analytical_solution}), we derive a closed-form solution for the parameters of the offset model in the case of a binary covariate.
This closed-form can be used to characterize the bias in offset solutions, though the resulting formula is opaque.
Therefore, we use the closed form for an extensive empirical evaluation.
First we investigate the relative efficiency of the \gls{mcm} compared to unconstrained estimation when both are consistent.
Then we investigate the \gls{pehe} of the offset model under a) non-collapsibility but no unobserved confounding and b) unobserved confounding but no non-collapsibility.
Finally, we compare the \gls{pehe} of three different \gls{cate} approximators: the offset model with marginal odds-ratio $\gamma$, \gls{mcm} and \gls{crmcm} and study when these models have better \gls{pehe} than the \gls{ateb} in a large experimental grid with a binary covariate and an unobserved confounder $U$.
\paragraph{Implementation.}
For implementing \gls{mcm} and \gls{crmcm}, we optimize the constrained objective \ref{eq:lagrangian} by specifying a small value $\lambda_0 = 0.01$ and optimizing the resulting unconstrained objective with a LBFGS optimizer.
After convergence of the unconstrained optimizer yielding model estimate $\hat{g}$, we evaluate the constraint $(M(\hat{g}) - \gamma)^2< \epsilon = 10^{-4}$.
If this is not satisfied, for the next iteration $i=1,2,...$ we increase $\lambda$ such that $\lambda_i = 10 * \lambda_{i-1}$ and repeat until the constraint is satisfied.
We relied on the JAX \citep{bradbury_jax_2018}, JAXOpt \citep{blondel_efficient_2022} and NumPyro \citep{phan_composable_2019} python libraries for implementing the experiments.
The code to reproduce all experiments is publicly available at www.github.com/vanamsterdam/binarymcm (DOI: 10.5281/zenodo.8144896).

\subsection{Relative efficiency of the marginally constrained estimator}
\label{sec:exp_efficiency}

Under the assumptions of strong ignorability, positivity and consistency and with a known marginal odds-ratio $\gamma$, we have at least two consistent estimators of the \gls{cate}: the unconstrained maximum-likelihood estimator and the \acrfull{mcm}.
In order to test the relative efficiency of the constrained versus the unconstrained estimator, we define an experimental grid with a binary covariate $X$ and data generating mechanism $\sigma^{-1}(\Pr(Y_t=1|x)) = \beta_0 + \beta_t t + \beta_x x$.
The parameter grid is given in Table~\ref{tab:grid_efficiency}.

\begin{table}[htpb]
	\centering
	\caption{Grid for experiment on relative efficiency of the marginally constrained versus the unconstrained estimator}
	\label{tab:grid_efficiency}
	\begin{tabular}{lll}
	parameter & definition                                                  & values                               \\ \hline
	$q_x$     & $\Pr(X=1)$                                                  & 0.5                                  \\
	$\beta_0$ & $\sigma^{-1}(\Pr(Y_0=1|X=0))$                               & $\sigma^{-1}(0.15,0.5)$ \\
	$\beta_t$ & $\sigma^{-1}(\Pr(Y_1=1|X)) - \sigma^{-1}(\Pr(Y_0=1|X))$     & 1                                    \\
	$\beta_x$ & $\sigma^{-1}(\Pr(Y_t=1|X=1)) - \sigma^{-1}(\Pr(Y_t=1|X=0))$ & log(1/5, 1/2, 1, 2, 5)               \\
	$q_t$     & $\Pr(T=1)$                                              & 0.5                                  \\
	$\eta_x$  & $\sigma^{-1}(\Pr(T=1|X=1)) - \sigma^{-1}(\Pr(T=1|X=0))$     & log(1/5, 1/2, 1, 2, 5)               \\ \hline
	$n$       & sample size  						& 50, 75, 100, 150, 200, 325, 400, 500, 750
	\end{tabular}
\end{table}

In this setup, a non-parametric model family for $\Pr(Y_t=1|X)$ is

\begin{equation*}
	\mathcal{G}= \left\{ g: \{0,1\}^2 \to [0,1], g(t,x) = \sigma(b_0 + b_t t + b_x x + b_{tx} tx), b_0,b_t,b_x,b_{tx} \in \mathbb{R} \right\}
\end{equation*}

Given this model family, sample log-likelihood estimator $L_n$, sample marginalizer $M_n$, and marginal log odds-ratio $\gamma$ (calculated from the parameters of the data generating mechanism), we compare the constrained and unconstrained estimator.
For each of the possible combinations of parameter values in Table \ref{tab:grid_efficiency}, we generated 250 datasets and applied both estimators.
Each fitted model produces an estimate of \gls{cate}$(X=0)$ and \gls{cate}$(X=1)$.
We construct 95\% confidence bounds for these estimated \glspl{cate} by taking the 2.5\% and 97.5\% percentile-values over the 250 repetitions for each simulation setting, sample size, estimator and $X$.
The width of each confidence bound (len(CI)) is a measure of statistical uncertainty that is relevant for treatment decision making.
As expected by asymptotic theory for asymptotically linear normal estimators, we found empirically that $\log (\text{len(CI)})$ was approximately linear in the logarithm of the sample size for each parameter setting and estimator, see Figures \ref{fig:app_ci_vs_n}, \ref{fig:app_log_n_vs_log_ci} in the Appendix.
We fit models to the experimental results to summarize the relative efficiency.
Specifically, denoting $m=1$ to indicate the unconstrained estimator and $m=0$ for the constrained estimator, we fit linear regression models of the form $\log n = w_0 + w_l \log(\text{len(CI)}) + w_m m + \epsilon$ to the experimental results of each parameter combination.
We took len(CI) to be the average of the logarithm of the confidence bound length of \gls{cate}$(X=1)$ and \gls{cate}$(X=0)$ in each experiment.

The linear models provided a good fit of the experimental results with an adjusted $R^2>0.975$ for all of the combinations of simulation parameters.
The parameter $w_m$ in this linear model has the following interpretation: if sample size $n_c$ reaches a certain length of the confidence bound with the constrained estimator,
to reach the same length of the confidence bound with the unconstrained method, $n_c e^{w_m}$ samples are needed, meaning a $100(e^{w_m} - 1)$\% increase in sample size.
Across all simulation settings, we find that unconstrained estimation requires at least 71\% and at most 106\% more samples, meaning that to achieve equal-width confidence bounds in the \gls{cate} estimation one needs at least 71\% more patients with the unconstrained estimator then when using the constrained estimator \gls{mcm}.

\subsection{The effect of non-collapsibility on the offset method}
\label{sec:exp_collapsibility}

When the offset method is applied using the \emph{marginal} log odds-ratio $\gamma$ in the place of \emph{conditional} log odds-ratio $\beta_t$, the offset model is inconsistent and has non-zero \gls{pehe}.
We conducted a numerical experiment to evaluate the \gls{pehe} of the offset model compared with the \gls{ateb} and \gls{crmcm} with a single binary covariate $X$ with varying effect on $Y$.
The data-generating mechanism for this experiment is $\sigma^{-1}(\Pr(Y_t=1|x)) = \beta_0 + \beta_t t + \beta_x x$, with $\beta_0 = 0, \beta_t = 1, 0 \leq \beta_x \leq \log 15 = 2.71$.
Furthermore, $\Pr(X=1) = \Pr(T=1) =0.5$.

\begin{figure}[htpb]
	\centering
	\includegraphics[width=0.8\textwidth]{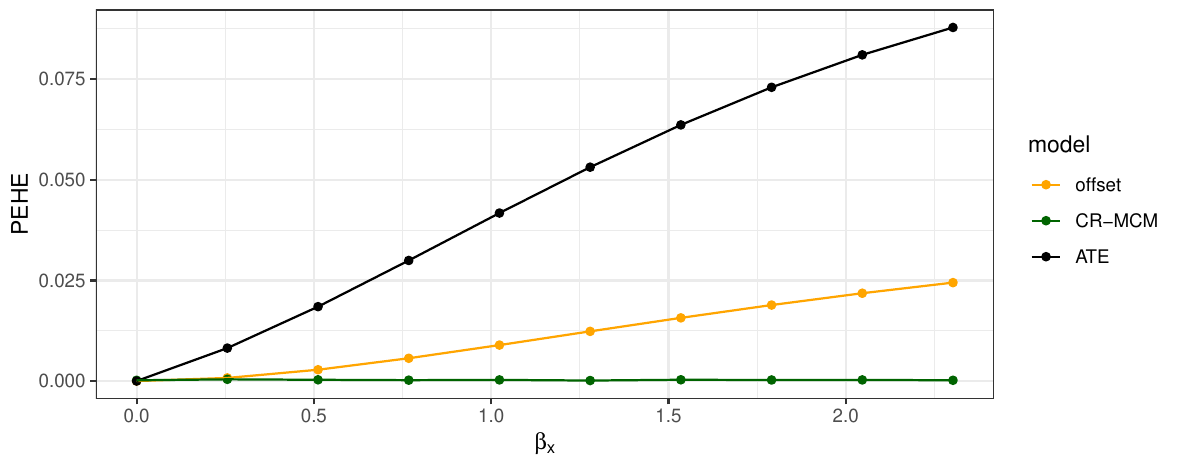}
	\caption{Experiment on the effect of non-collapsibility on the offset model when the \emph{marginal} log odds-ratio $\gamma$ is used instead of the \emph{conditional} log odds-ratio $\beta_t$, for varying effects of $X$ on $Y$.
		When the outcome depends more on covariate $x$ (i.e. bigger $\beta_x$) the difference between the \emph{marginal} odds-ratio and \emph{conditional} odds-ratio becomes bigger due to non-collapsibility of the odds-ratio, leading to worse \gls{cate} approximation error for the offset model.
		The \gls{crmcm} model uses the marginal odds-ratio as a constraint and remains consistent.
		\gls{crmcm}: \acrlong{crmcm}, \gls{ate}: \acrlong{ate}
	}
	\label{fig:collapsibility}
\end{figure}

In Figure \ref{fig:collapsibility} we see that when $\beta_x$ increases, the \gls{pehe} of the offset model with $\gamma$ as an offset increases as expected due to the increasing difference between $\gamma$ and $\beta_t$.
However, the \gls{pehe} of the \gls{ateb} increases faster.
Finally, we note that the \gls{crmcm} remains consistent as by Theorem \ref{th:constr}.

\subsection{The effect of unobserved confounding on the offset method}
\label{sec:exp_confounding}

When the strong ignorability does not hold, for example due to the presence of an unobserved confounder $U$, Example 1 (section \ref{sec:example1}) demonstrates that the offset model is an inconsistent estimator of $\Pr(Y_t=1)$, even if the conditional log odds-ratio $\beta_t$ is known.
As motivated before, it may still be justifiable to use the offset method when it has better \gls{pehe} than the \gls{ateb}.
To study the effect of unobserved confounding on the \gls{pehe} of the offset method, we conducted an experiment with a binary covariate $X$, binary unobserved confounder $U$ and in this case a known \emph{conditional} log odds-ratio $\beta_t$.
We note that this $\beta_t$ is generally not available from \glspl{rct}, but we use it in this experiment to separate out the effect of unobserved confounding from non-collapsibility.
The data-generating mechanism for this experiment is:

\begin{itemize}
	\item $\sigma^{-1}(\Pr(Y_t=1|X=x,U=u)) = \alpha_0 + \alpha_t t + \alpha_x x + \alpha_{tx} tx + \alpha_u u $
	\item $\Pr(U|T=t)$ such that $\sigma^{-1}(\Pr(U=1|T=1)) - \sigma^{-1}(\Pr(U=1|T=0)) = \gamma_u = \alpha_u$.
\end{itemize}

The parameter ranges were $\alpha_0 = \sigma^{-1}(0.05), \alpha_t = 1, \alpha_x \in \log(1, 2, 3, 4, 5), \alpha_u \in \log(1,2 ,5)$.
We set the marginals $\Pr(U=1)=\Pr(X=1)=\Pr(T=1)=0.5$.
Parameter $\alpha_x$ controls the effect of $X$ on $Y$ and $\alpha_u$ determines the amount of unobserved confounding (both through $U \to Y$ and $U \to T$).
Note that if the offset assumption in Equation \ref{eq:offset_assumption} is used, it is assumed to hold conditional on the observed covariate $X$, i.e. for distribution $\Pr(Y_t|X)$, not for $\Pr(Y_t|X,U)$.
Therefore, given values for $\alpha_0, \alpha_t, \alpha_x, \alpha_u, \Pr(U|T=t)$, the parameter $\alpha_{tx}$ was determined such that the offset assumption was valid for $\Pr(Y_t|X)$.
To implement \gls{crmcm} the \emph{marginal} log odds-ratio for treatment $\gamma$ was calculated from the experiment parameters.

\begin{figure}[htpb]
	\centering
	\includegraphics[width=1\textwidth]{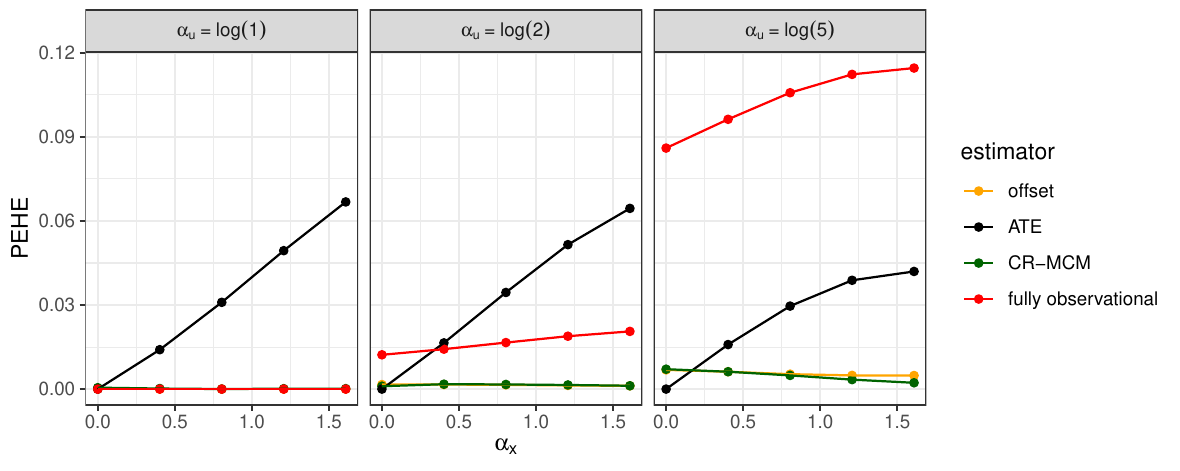}
	\caption{Experiment on the effect of unobserved confounding on the offset estimator when the \emph{conditional} log odds-ratio $\beta_t$ is known but there is an unobserved confounder, for varying effects of $X$ on $Y$.
		$\alpha_x$ determines the effect of $X$ on the outcome $Y$, $\alpha_u$ determines the amount of unobserved confounding.
		When $\alpha_x$ gets bigger, the average treatment effect becomes a bad approximator for the \gls{cate}.
		When unobserved confounding increases, the fully observational baseline becomes a bad approximator of the \gls{cate} as expected.
		Both the offset model and the \gls{crmcm} perform well, although the offset model requires knowledge of the conditional odds-ratio which is generally not available, while \gls{crmcm} only requires the marginal odds-ratio which is generally reported in \glspl{rct}.
		\gls{crmcm}: \acrlong{crmcm}; \gls{ate}: \acrlong{ate}
	}
	\label{fig:confounding}
\end{figure}

On this experimental grid four different models are compared: the \gls{ateb}, the offset method with conditional odds-ratio, \gls{crmcm} and a `fully-observational' estimator, i.e. the non-parametric maximum likelihood estimator of $q(Y=1|T,X)$ where $q$ denotes the observational distribution.
The results are presented in Figure \ref{fig:confounding}.
As expected, the \gls{pehe} of the fully-observational estimator is very sensitive to increasing amounts of unobserved confounding.
The \gls{pehe} of the \gls{ateb} increases when the variance of the \gls{cate} increases, which in this experiment is determined by $\alpha_x$.
While the \gls{pehe} for the offset method increases when unobserved confounding increases, it still has better \gls{pehe} than the \gls{ateb} when $\alpha_x \neq 0$.
Finally, \gls{crmcm} is the best performing model overall even though, in contrast with the offset model, does not require knowledge of the conditional odds-ratio $\beta_t$ which is generally not available from \glspl{rct}.

\pagebreak

\subsection{Marginally constrained CATE approximation in the presence of unobserved confounding and non-collapsibility}

Finally, we investigate whether different \gls{cate} approximators lead to better \gls{pehe} than the \gls{ateb} in an experimental grid covering the space of configurations for a binary covariate $X$ and binary unobserved confounder $U$.
The parameter grid and values are presented in Table \ref{tab:fullgrid} and Table \ref{tab:fullgrid_prmvals}.

\begin{table}[htpb]
	\centering
	\caption{Grid for experiment on \gls{pehe} of \gls{cate} approximators versus \gls{ateb}, definition of parameters}
	\label{tab:fullgrid}
	\begin{tabular}{lll}
	distribution & parameters \\ \hline
	$\Pr(Y_t|X)$ & $\Pr(Y=1|t,x,u) = \sigma \left (\alpha_0 + \alpha_t t + \alpha_x x + \alpha_u u + \alpha_{tx} tx + \alpha_{tu} tu + \alpha_{xu} xu + \alpha_{txu} txu \right) $  \\
	$\Pr(U|T)$ & $\gamma_u = \sigma^{-1}(\Pr(U=1|T=1)) - \sigma^{-1}(\Pr(U=1|T=0))$                               \\
	$\Pr(U)$ & $\Pr(U=1)$     \\
	$\Pr(X)$ & $\Pr(X=1)$     \\
	$\Pr(T)$ & $\Pr(T=1)$     
	\end{tabular}
\end{table}

\begin{table}[htpb]
	\centering
	\caption{Grid for experiment on \gls{pehe} of \gls{crmcm} versus \gls{ateb}, parameter values; $^{\dagger} \alpha_{tx}$ takes these values only when enforce offset = False}
	\label{tab:fullgrid_prmvals}
	\begin{tabular}{lll}
	values & parameters \\ \hline
	logit$(0.025,0.15, 0.5, 0.85)$ & $\alpha_0$ \\
	$0, 0.402, 0.804, 0.121, 1.161 $ & $\alpha_x,\alpha_u,\alpha_{tu},\alpha_{xu},\alpha_{txu},\alpha_{tx}^{\dagger} $  \\
	$0.402, 0.804, 0.121, 1.161 $ & $\alpha_t$  \\
	$-1.161, -1.207, -0.804, -0.402, 0, 0.402, 0.804, 1.207, 1.161 $ & $\gamma_u$  \\
	$0.2, 0.5 $ & $\Pr(U=1), \Pr(X=1), \Pr(T=1)$     \\
	True, False & enforce offset 
	\end{tabular}
\end{table}

For the parameter $\alpha_{tx}$ we took one of two approaches depending on `enforce offset': either one of the values listed in Table \ref{tab:fullgrid_prmvals} when `enforce offset' = False, or the value that makes the offset assumption satisfied in $\Pr(Y_t=1|X)$ when `enforce offset' = True.

To summarize the results we formulated a criterion based on the variance of the \gls{cate} as when the \gls{cate} is constant, the \gls{pehe} of the \gls{ateb} is 0 and no model can improve on that.
The higher the variance of the \gls{cate}, the more \gls{cate} models can potentially improve on the \gls{ateb}.
We group experiments by setting a threshold $\tau$ for the variance of the \gls{cate}.
To parameterize $\tau$ we turn to the scale of the \glspl{cate}, under the assumption that $\Pr(X=1) = 0.5$, the simplest version of the experiment.
This makes $\tau$ a function of only the \glspl{cate} which makes it more interpretable.
To express $\tau$ on the \gls{cate} scale of the we introduce $\delta$ as the difference between the \glspl{cate} depending on $X$:

\begin{equation*}
	\delta := | \text{CATE}(X=1) - \text{CATE}(X=0) |
\end{equation*}

Given $\delta$ and $\Pr(X=1) = 0.5$ we can calculate $\tau$ as:

\begin{align*}
	\tau &= \text{VAR(CATE)} \\
	     &= \EX_x \left[ \left( \text{CATE}(x) - \EX [ \text{CATE}(x)] \right)^2 \right] \\
	     &= \Pr(X=0) \left( \text{CATE}(X=0)  - \left(\Pr(X=0) \text{CATE}(X=0) + \Pr(X=1) \text{CATE}(X=1) \right) \right)^2 \\
	     &+ \Pr(X=1) \left( \text{CATE}(X=1)  - \left(\Pr(X=0) \text{CATE}(X=0) + \Pr(X=1) \text{CATE}(X=1) \right) \right)^2 \\
	     &= 0.5 \left( 0.5 \text{CATE}(X=0)  - 0.5  \text{CATE}(X=1) \right)^2 \\
	     &+ 0.5 \left( 0.5 \text{CATE}(X=1)  - 0.5  \text{CATE}(X=0) \right)^2 \\
	     &= 0.5 * 0.25 * \left( \delta^2 + \delta^2  \right) \\
	     &= 0.25 \delta^2
\end{align*}

For each $\tau$ we subsetted the experiments for which $\text{VAR(CATE)} > \tau$ and we evaluated in what percentage of experiments the \gls{cate} approximators performed better than the \gls{ateb}.
Furthermore, we split the results for wheter the offset assumption did or did not hold.

We took $\delta \in \{0.01, 0.05\}$ corresponding to relatively small differences in \glspl{cate}.
As shown in Table \ref{tab:fullgrid_results_offset}, when $\delta$ increases, for all \gls{cate} approximators the fraction of experiments with improved \gls{pehe} compared to the \gls{ateb} increases.
The offset model with marginal odds-ratio and the \gls{crmcm} improve relative to \gls{mcm} when the offset assumption is indeed satisfied.
Taking the experiments in which the offset assumption is satisfied together with those in which it is not,
\gls{mcm} performs best with $>91\%$ of experiments having better \gls{pehe} than the \gls{ateb} when $\delta > 0$, $>97\%$ when $\delta > 0.01$ and $>99\%$ when $\delta > 0.05$.
Even though the offset model with marginal odds-ratio is the worst performing model in the comparison, it still leads to better \gls{pehe} estimation than the \gls{ateb} when $\delta > 0.05$ in $>80\%$ of the experiments when the offset assumption does not hold, and $>98\%$ of experiments where the offset assumption does hold.
This indicates that the \gls{crate} models in clinical use \citep{candido_dos_reis_updated_2017,ravdin_computer_2001} might provide value for individualized treatment decision making even if they were developed in data with unobserved confounding.

\begin{table}[ht]
	\centering
	   \caption{PEHE of different CATE approximators for varying values of $\delta:=| \text{CATE}(X=1) - \text{CATE}(X=0) |$.
		   The experiments were filtered such that VAR(CATE)$>\frac{1}{4}\delta^2$, which is the variance of the \gls{cate} for that value of $\delta$ when $\Pr(X=1)=0.5$.
		   The numbers in columns offset, \gls{crmcm}, \gls{mcm} denote in what percentage of the experiments that model had better \gls{pehe} than the \gls{ateb}.
			offset: offset approach with \emph{marginal} odds-ratio used in the place of the conditional odds-ratio,
			\gls{crmcm}: \acrlong{crmcm},
			\gls{mcm}: \acrlong{mcm},
			N: number of experiments satisfying the VAR(CATE) and `offset satisfied' requirement}
	\label{tab:fullgrid_results_offset}
	\begin{tabular}{rlrrrr}
	  \hline
	$\delta$ & offset satisfied & offset & CR-MCM & MCM & N \\ 
	  \hline
	0.00 & False & 0.68 & 0.70 & \textbf{0.96} & 16455472 \\ 
	0.00 & True & 0.70 & \textbf{0.77} & 0.69 & 3282598 \\ 
	0.01 & False & 0.71 & 0.73 & \textbf{0.98} & 15777278 \\ 
	0.01 & True & 0.90 & \textbf{0.97} & 0.89 & 2497022 \\ 
	0.05 & False & 0.80 & 0.81 & \textbf{1.00} & 13193906 \\ 
	0.05 & True & 0.98 & \textbf{0.99} & 0.98 & 1295294 \\ 
	   \hline
	\end{tabular}
\end{table}

\section{Discussion}

We introduced \acrfullpl{mcm} that make use of a known marginal treatment effect to approximate the \gls{cate} from observational data.
Under the standard causal inference assumptions and some regularity conditions \glspl{mcm} are consistent estimators of the \gls{cate} and in our experiments they are also more efficient than unconstrained estimation.
Next, in the presence of unobserved confounding, we showed that the offset method does not provide consistent \gls{cate} estimates for binary outcomes.
We find that \glspl{mcm} tend to have better \gls{pehe} than offset models, and both models have better \gls{pehe} than the \gls{ateb} in almost all settings in the case of a binary covariate, as long as the variance in the \gls{cate} is greater than a minimal threshold. Further, when the constant relative effect assumption holds, \glspl{crmcm} are even better. 

An important question for the offset model and the \gls{crmcm} is when it is valid to assume that the relative treatment effect is indeed constant conditional on the patient features.
There is some evidence from meta-analyses that treatment effect estimates on a relative scale are more stable across different \glspl{rct} than treatment effects on an absolute scale \citep{engels_heterogeneity_2000}.
However, in some settings there may be clear indications for differences in treatment effect on a relative scale.
For instance, breast cancer patients respond better to estrogen receptor modulator tamoxifen when they have an estrogen-sensitive tumor \cite{early_breast_cancer_trialists_collaborative_group_tamoxifen_1998}.
When the difference in relative treatment effect is known, this difference could be accounted for accordingly in \gls{mcm} and offset models.
The constant odds-ratio for treatment remains a strong parametric assumption, though in our experiments we found that the offset model and \gls{crmcm} tend to have better \gls{pehe} than the \gls{ateb} even if the offset assumption does not hold,
as long as the variance in the \gls{cate} is greater than a minimal threshold.

Instead of only using a single treatment effect estimate from prior \glspl{rct}, recent work has studied combining observational data and data from \glspl{rct} for \gls{cate} estimation \citep{rosenman_combining_2020,ilse_combining_2022}.
Under relatively mild assumptions, estimates from combined datasets yield more efficient estimates of \glspl{cate} than using \gls{rct} data alone.
However, these methods require access to the individual-patient data from the \gls{rct}, whereas \glspl{mcm} only rely on a single effect estimate from \glspl{rct} which is usually available from published \gls{rct} results.
Gaining access to individual-patient data from \glspl{rct} is often challenging due to data-access restrictions.

We only studied settings with a single binary covariate in our experiments.
Future work could experiment with higher dimensional, mixed-type covariates.
In higher dimensions, the constraint on the implied marginal odds-ratio restricts a lower fraction of the degrees of freedom.
It is unknown whether the constraint will help attain better \gls{cate} approximation in the presence of unobserved confounding with higher dimensional parameter spaces and the relative efficiency gain of using the constraint may be smaller. 
Machine learning methods that learn lower-dimensional representations of covariate distributions that still preserve the information relevant for the untreated risk might help restore the efficiency.

Future work could extend our experiments to relative treatment effect estimates for time-to-event outcomes such as the hazard-ratio, or to the settings of time-varying treatments and confounding.
Furthermore, we assumed that the \gls{rct} gives an unbiased estimate of $\gamma$ with infinite precision.
In practice, \glspl{rct} are often conducted in non-random samples of the population which may result in different covariate distributions $p_{\text{rct}}(X) \neq q_{\text{obs}}(X)$.
If $\Pr(Y_t|X)$ is transportable from the \gls{rct} to the observational data,
the marginal odds-ratio from the \gls{rct} will be different from the implied marginal odds-ratio of $\Pr(Y_t|X)$ when calculated in the observational distribution,
because $\Pr(Y_t|X)$ is marginalized over a different distribution of $X$.
However, if $p_{\text{rct}}(X)$ and $q_{\text{obs}}(X)$ are known and appropriate sampling weights exist the constraint on the marginal odds-ratio may be applied using these weights.

Bayesian extensions of \glspl{mcm} can be investigated to account for uncertainty in marginal odds-ratio estimates from \glspl{rct}.
Finally, finite-sample characteristics of the estimator for the implied marginal odds-ratio in terms of bias and variance could be studied further.

\section*{Acknowledgments}
We are grateful to Nan van Geloven for providing feedback on an earlier version of this manuscript.

\section*{Funding}
RR was partly supported by NIH/NHLBI Award R01HL148248, NSF CAREER Award 2145542 and by NSF Award 1922658 NRT-HDR: FUTURE Foundations, Translation, and Responsibility for Data Science.
WA reports no specific funding for this project.
WA was employed by Babylon Health Inc during this research project but now works at the University Medical Center Utrecht, the Netherlands.

\section*{Author contributions}
All authors have accepted responsibility for the entire content of this manuscript and approved its submission.

\section*{Conflict of interest}
Authors state no conflict of interest.

\section*{Data availability}

The code to reproduce all experiments is publicly available at www.github.com/vanamsterdam/binarymcm (DOI: 10.5281/zenodo.8144896)

{
\small
\bibliography{mybib}

\begin{thebibliography}{10}

\bibitem{murray_patients_2018}
Murray EJ, Caniglia EC, Swanson SA, Hernández-Díaz S, Hernán MA.
\newblock Patients and investigators prefer measures of absolute risk in
  subgroups for pragmatic randomized trials.
\newblock Journal of Clinical Epidemiology. 2018 Nov;103:10--21.

\bibitem{pearl_causal_2009}
Causal {Diagrams} and the {Identification} of {Causal} {Effects}.
\newblock In: Pearl J, editor. Causality. Cambridge: Cambridge University
  Press; 2009. p. 65--106.
\newblock Available from:
  \url{https://www.cambridge.org/core/books/causality/causal-diagrams-and-the-identification-of-causal-effects/D9AE074727C3AC9AFE9F0CD4C7A506B5}.

\bibitem{furie_two-year_2020}
Furie R, Rovin BH, Houssiau F, Malvar A, Teng YKO, Contreras G, et~al.
\newblock Two-{Year}, {Randomized}, {Controlled} {Trial} of {Belimumab} in
  {Lupus} {Nephritis}.
\newblock The New England Journal of Medicine. 2020 Sep;383(12):1117--1128.

\bibitem{lean_primary_2018}
Lean ME, Leslie WS, Barnes AC, Brosnahan N, Thom G, McCombie L, et~al.
\newblock Primary care-led weight management for remission of type 2 diabetes
  ({DiRECT}): an open-label, cluster-randomised trial.
\newblock Lancet (London, England). 2018 Feb;391(10120):541--551.

\bibitem{candido_dos_reis_updated_2017}
Candido~dos Reis FJ, Wishart GC, Dicks EM, Greenberg D, Rashbass J, Schmidt MK,
  et~al.
\newblock An updated {PREDICT} breast cancer prognostication and treatment
  benefit prediction model with independent validation.
\newblock Breast Cancer Research. 2017 Dec;19(1):58.
\newblock 80 citations (Crossref) [2021-08-06].
\newblock Available from:
  \url{http://breast-cancer-research.biomedcentral.com/articles/10.1186/s13058-017-0852-3}.

\bibitem{ravdin_computer_2001}
Ravdin PM, Siminoff LA, Davis GJ, Mercer MB, Hewlett J, Gerson N, et~al.
\newblock Computer {Program} to {Assist} in {Making} {Decisions} {About}
  {Adjuvant} {Therapy} for {Women} {With} {Early} {Breast} {Cancer}.
\newblock Journal of Clinical Oncology. 2001 Feb;19(4):980--991.
\newblock 679 citations (Crossref) [2021-08-06].
\newblock Available from:
  \url{http://ascopubs.org/doi/10.1200/JCO.2001.19.4.980}.

\bibitem{alaa_machine_2021}
Alaa AM, Gurdasani D, Harris AL, Rashbass J, van~der Schaar M.
\newblock Machine learning to guide the use of adjuvant therapies for breast
  cancer.
\newblock Nature Machine Intelligence. 2021 Aug;3(8):716--726.
\newblock Bandiera\_abtest: a Cg\_type: Nature Research Journals Number: 8
  Primary\_atype: Research Publisher: Nature Publishing Group Subject\_term:
  Breast cancer;Prognosis Subject\_term\_id: breast-cancer;prognosis.
\newblock Available from:
  \url{https://www.nature.com/articles/s42256-021-00353-8}.

\bibitem{xu_prediction_2021}
Xu Z, Arnold M, Stevens D, Kaptoge S, Pennells L, Sweeting MJ, et~al.
\newblock Prediction of {Cardiovascular} {Disease} {Risk} {Accounting} for
  {Future} {Initiation} of {Statin} {Treatment}.
\newblock American Journal of Epidemiology. 2021 Feb:kwab031.
\newblock Available from:
  \url{https://academic.oup.com/aje/advance-article/doi/10.1093/aje/kwab031/6140872}.

\bibitem{cardoso_early_2019}
Cardoso F, Kyriakides S, Ohno S, Penault-Llorca F, Poortmans P, Rubio IT,
  et~al.
\newblock Early breast cancer: {ESMO} {Clinical} {Practice} {Guidelines} for
  diagnosis, treatment and follow-up.
\newblock Annals of Oncology. 2019 Aug;30(8):1194--1220.
\newblock Available from:
  \url{https://linkinghub.elsevier.com/retrieve/pii/S0923753419312876}.

\bibitem{gradishar_nccn_2021}
Gradishar WJ. {NCCN} {Breast} {Cancer} {Guideline}, {Version} 5.2021; 2021.
\newblock Available from:
  \url{https://www.nccn.org/professionals/physician_gls/pdf/breast-2.pdf}.

\bibitem{hill_bayesian_2011}
Hill JL.
\newblock Bayesian {Nonparametric} {Modeling} for {Causal} {Inference}.
\newblock Journal of Computational and Graphical Statistics. 2011
  Jan;20(1):217--240.
\newblock Available from:
  \url{http://www.tandfonline.com/doi/abs/10.1198/jcgs.2010.08162}.

\bibitem{suverein_early_2023}
Suverein MM, Delnoij TSR, Lorusso R, Brandon Bravo~Bruinsma GJ, Otterspoor L,
  Elzo~Kraemer CV, et~al.
\newblock Early {Extracorporeal} {CPR} for {Refractory} {Out}-of-{Hospital}
  {Cardiac} {Arrest}.
\newblock New England Journal of Medicine. 2023 Jan;388(4):299--309.
\newblock Available from: \url{http://www.nejm.org/doi/10.1056/NEJMoa2204511}.

\bibitem{hill_methylprednisolone_2022}
Hill KD, Kannankeril PJ, Jacobs JP, Baldwin HS, Jacobs ML, O’Brien SM, et~al.
\newblock Methylprednisolone for {Heart} {Surgery} in {Infants} — {A}
  {Randomized}, {Controlled} {Trial}.
\newblock New England Journal of Medicine. 2022 Dec;387(23):2138--2149.
\newblock Available from: \url{http://www.nejm.org/doi/10.1056/NEJMoa2212667}.

\bibitem{weghuber_once-weekly_2022}
Weghuber D, Barrett T, Barrientos-Pérez M, Gies I, Hesse D, Jeppesen OK,
  et~al.
\newblock Once-{Weekly} {Semaglutide} in {Adolescents} with {Obesity}.
\newblock New England Journal of Medicine. 2022 Dec;387(24):2245--2257.
\newblock Available from: \url{http://www.nejm.org/doi/10.1056/NEJMoa2208601}.

\bibitem{tchetgen_tchetgen_doubly_2010}
Tchetgen~Tchetgen EJ, Robins JM, Rotnitzky A.
\newblock On doubly robust estimation in a semiparametric odds ratio model.
\newblock Biometrika. 2010 Mar;97(1):171--180.
\newblock Available from:
  \url{https://www.ncbi.nlm.nih.gov/pmc/articles/PMC3412601/}.

\bibitem{tchetgen_tchetgen_double-robust_2011}
Tchetgen~Tchetgen EJ, Rotnitzky A.
\newblock Double-robust estimation of an exposure-outcome odds ratio adjusting
  for confounding in cohort and case-control studies.
\newblock Statistics in Medicine. 2011 Feb;30(4):335--347.

\bibitem{watson_practitioners_2007}
Watson T.
\newblock Practitioner's {Guide} to {Generalized} {Linear} {Models}.
\newblock Towers Watson; 2007.

\bibitem{engels_heterogeneity_2000}
Engels EA, Schmid CH, Terrin N, Olkin I, Lau J.
\newblock Heterogeneity and statistical significance in meta-analysis: an
  empirical study of 125 meta-analyses.
\newblock Statistics in Medicine. 2000 Jul;19(13):1707--1728.

\bibitem{huitfeldt_collapsibility_2019}
Huitfeldt A, Stensrud MJ, Suzuki E.
\newblock On the collapsibility of measures of effect in the counterfactual
  causal framework.
\newblock Emerging Themes in Epidemiology. 2019 Jan;16(1):1.
\newblock Available from: \url{https://doi.org/10.1186/s12982-018-0083-9}.

\bibitem{didelez_logic_2022}
Didelez V, Stensrud MJ.
\newblock On the logic of collapsibility for causal effect measures.
\newblock Biometrical Journal. 2022;64(2):235--242.
\newblock \_eprint:
  https://onlinelibrary.wiley.com/doi/pdf/10.1002/bimj.202000305.
\newblock Available from:
  \url{https://onlinelibrary.wiley.com/doi/abs/10.1002/bimj.202000305}.

\bibitem{whittemore_collapsibility_1978}
Whittemore AS.
\newblock Collapsibility of {Multidimensional} {Contingency} {Tables}.
\newblock Journal of the Royal Statistical Society Series B (Methodological).
  1978;40(3):328--340.
\newblock Publisher: [Royal Statistical Society, Wiley].
\newblock Available from: \url{https://www.jstor.org/stable/2984697}.

\bibitem{daniel_making_2021}
Daniel R, Zhang J, Farewell D.
\newblock Making apples from oranges: {Comparing} noncollapsible effect
  estimators and their standard errors after adjustment for different covariate
  sets.
\newblock Biometrical Journal. 2021;63(3):528--557.
\newblock \_eprint:
  https://onlinelibrary.wiley.com/doi/pdf/10.1002/bimj.201900297.
\newblock Available from:
  \url{https://onlinelibrary.wiley.com/doi/abs/10.1002/bimj.201900297}.

\bibitem{hauck_consequence_1991}
Hauck WW, Neuhaus JM, Kalbfleisch JD, Anderson S.
\newblock A consequence of omitted covariates when estimating odds ratios.
\newblock Journal of Clinical Epidemiology. 1991 Jan;44(1):77--81.
\newblock Available from:
  \url{https://www.sciencedirect.com/science/article/pii/089543569190203L}.

\bibitem{pearl_overview_2009}
Pearl J.
\newblock Causal inference in statistics: {An} overview.
\newblock Statistics Surveys. 2009 Jan;3(none).
\newblock Available from:
  \url{https://projecteuclid.org/journals/statistics-surveys/volume-3/issue-none/Causal-inference-in-statistics-An-overview/10.1214/09-SS057.full}.

\bibitem{hernan_causal_nodate}
Hernan MA, Robins JM.
\newblock Causal {Inference}: {What} {If}. 2020.

\bibitem{wald_fitting_1940}
Wald A.
\newblock The {Fitting} of {Straight} {Lines} if {Both} {Variables} are
  {Subject} to {Error}.
\newblock The Annals of Mathematical Statistics. 1940 Sep;11(3):284--300.
\newblock Publisher: Institute of Mathematical Statistics.
\newblock Available from:
  \url{https://projecteuclid.org/journals/annals-of-mathematical-statistics/volume-11/issue-3/The-Fitting-of-Straight-Lines-if-Both-Variables-are-Subject/10.1214/aoms/1177731868.full}.

\bibitem{hartford2017deep}
Hartford J, Lewis G, Leyton-Brown K, Taddy M.
\newblock Deep IV: A flexible approach for counterfactual prediction.
\newblock In: International Conference on Machine Learning. PMLR; 2017. p.
  1414--1423.

\bibitem{puli2020general}
Puli A, Ranganath R.
\newblock General Control Functions for Causal Effect Estimation from IVs.
\newblock Advances in neural information processing systems.
  2020;33:8440--8451.

\bibitem{miao_identifying_2018}
Miao W, Geng Z, Tchetgen~Tchetgen EJ.
\newblock Identifying causal effects with proxy variables of an unmeasured
  confounder.
\newblock Biometrika. 2018 Dec;105(4):987--993.
\newblock Available from: \url{https://doi.org/10.1093/biomet/asy038}.

\bibitem{van_amsterdam_individual_2022}
van Amsterdam WAC, Verhoeff JJC, Harlianto NI, Bartholomeus GA, Puli AM,
  de~Jong PA, et~al.
\newblock Individual treatment effect estimation in the presence of unobserved
  confounding using proxies: a cohort study in stage {III} non-small cell lung
  cancer.
\newblock Scientific Reports. 2022 Apr;12(1):5848.
\newblock Number: 1 Publisher: Nature Publishing Group.
\newblock Available from:
  \url{https://www.nature.com/articles/s41598-022-09775-9}.

\bibitem{ilse_combining_2022}
Ilse M, Forré P, Welling M, Mooij JM.
\newblock Combining {Interventional} and {Observational} {Data} {Using}
  {Causal} {Reductions}.
\newblock arXiv:210304786 [cs, stat]. 2022 Jan.
\newblock ArXiv: 2103.04786.
\newblock Available from: \url{http://arxiv.org/abs/2103.04786}.

\bibitem{bradbury_jax_2018}
Bradbury J, Frostig R, Hawkins P, Johnson MJ, Leary C, Maclaurin D, et~al..
  {JAX}: composable transformations of {Python}+{NumPy} programs; 2018.
\newblock Available from: \url{http://github.com/google/jax}.

\bibitem{blondel_efficient_2022}
Blondel M, Berthet Q, Cuturi M, Frostig R, Hoyer S, Llinares-López F, et~al..
  Efficient and {Modular} {Implicit} {Differentiation}. arXiv; 2022.
\newblock ArXiv:2105.15183 [cs, math, stat].
\newblock Available from: \url{http://arxiv.org/abs/2105.15183}.

\bibitem{phan_composable_2019}
Phan D, Pradhan N, Jankowiak M.
\newblock Composable {Effects} for {Flexible} and {Accelerated} {Probabilistic}
  {Programming} in {NumPyro}.
\newblock arXiv preprint arXiv:191211554. 2019.

\bibitem{early_breast_cancer_trialists_collaborative_group_tamoxifen_1998}
Group EBCTC.
\newblock Tamoxifen for early breast cancer: an overview of the randomised
  trials.
\newblock The Lancet. 1998 May;351(9114):1451--1467.
\newblock Available from:
  \url{https://www.sciencedirect.com/science/article/pii/S0140673697114234}.

\bibitem{rosenman_combining_2020}
Rosenman E, Basse G, Owen A, Baiocchi M.
\newblock Combining {Observational} and {Experimental} {Datasets} {Using}
  {Shrinkage} {Estimators}.
\newblock arXiv:200206708 [math, stat]. 2020 May.
\newblock ArXiv: 2002.06708.
\newblock Available from: \url{http://arxiv.org/abs/2002.06708}.

\end{thebibliography}
}

\pagebreak

\appendix

\section{Appendix}

\subsection{Non-Collapsibility}
\label{app:nc}

\begin{figure}
	\centering
	\begin{tabular}{cc}
		\includegraphics[width=3in]{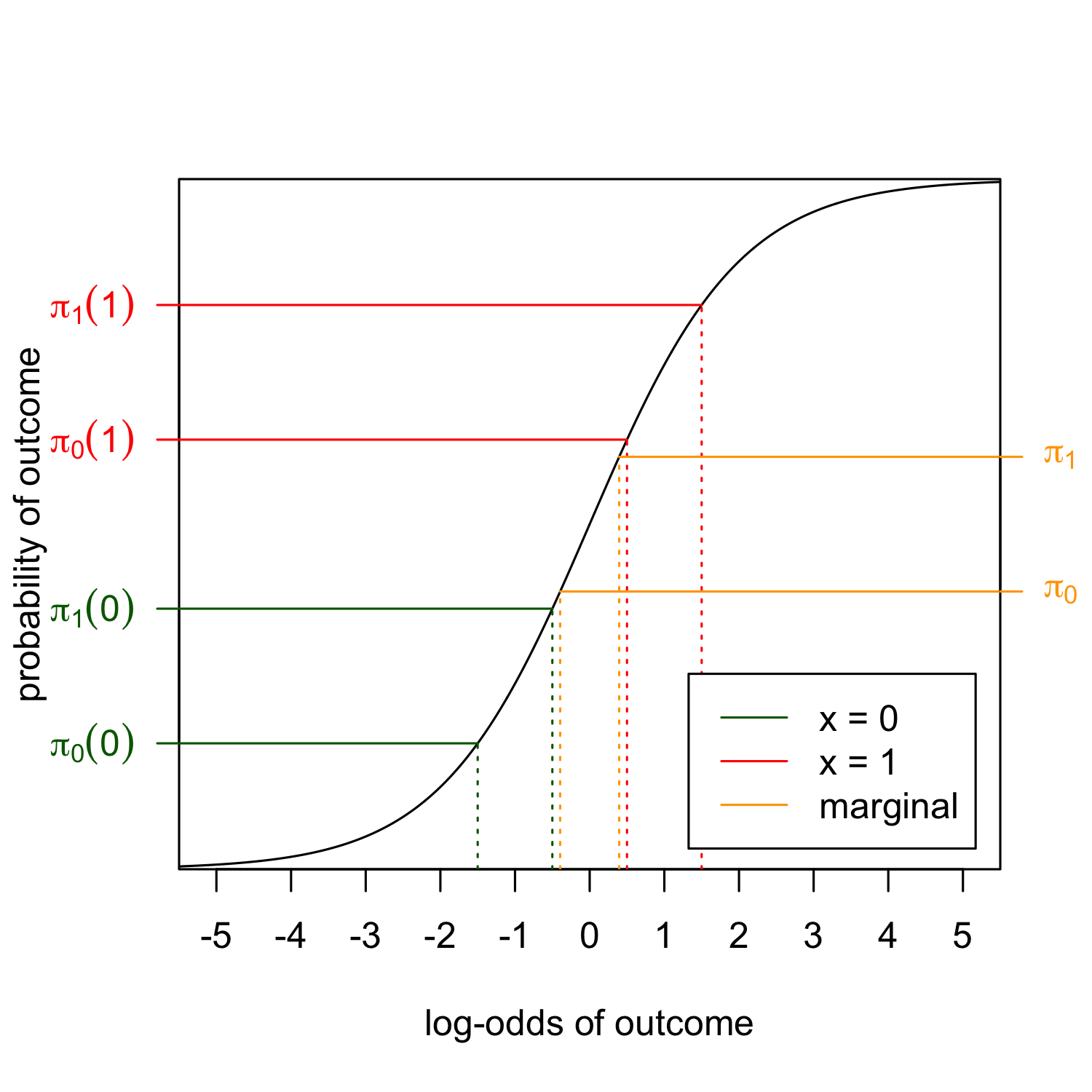} & \includegraphics[width=3in]{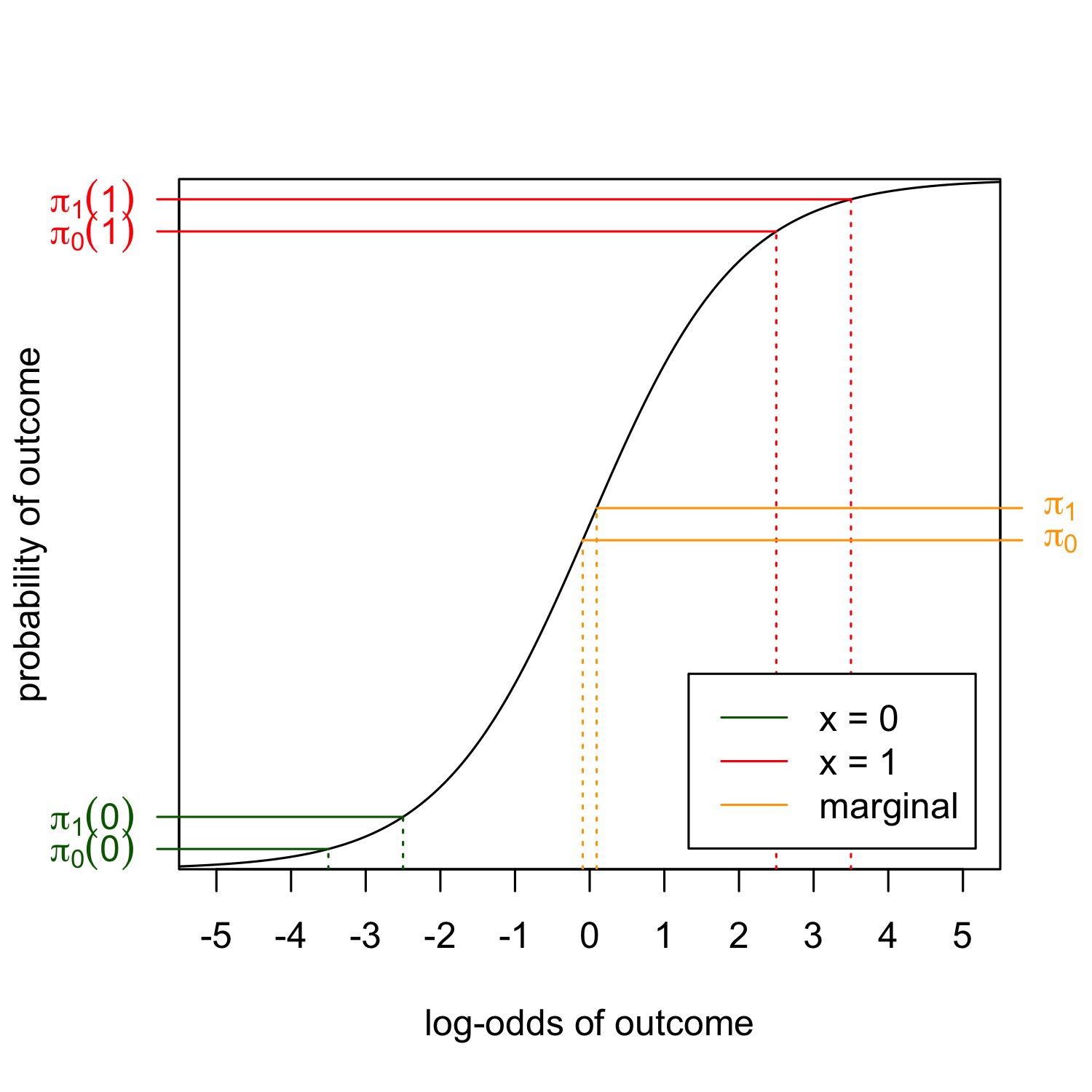}  \\
		\multicolumn{2}{c}{\includegraphics[height=3in]{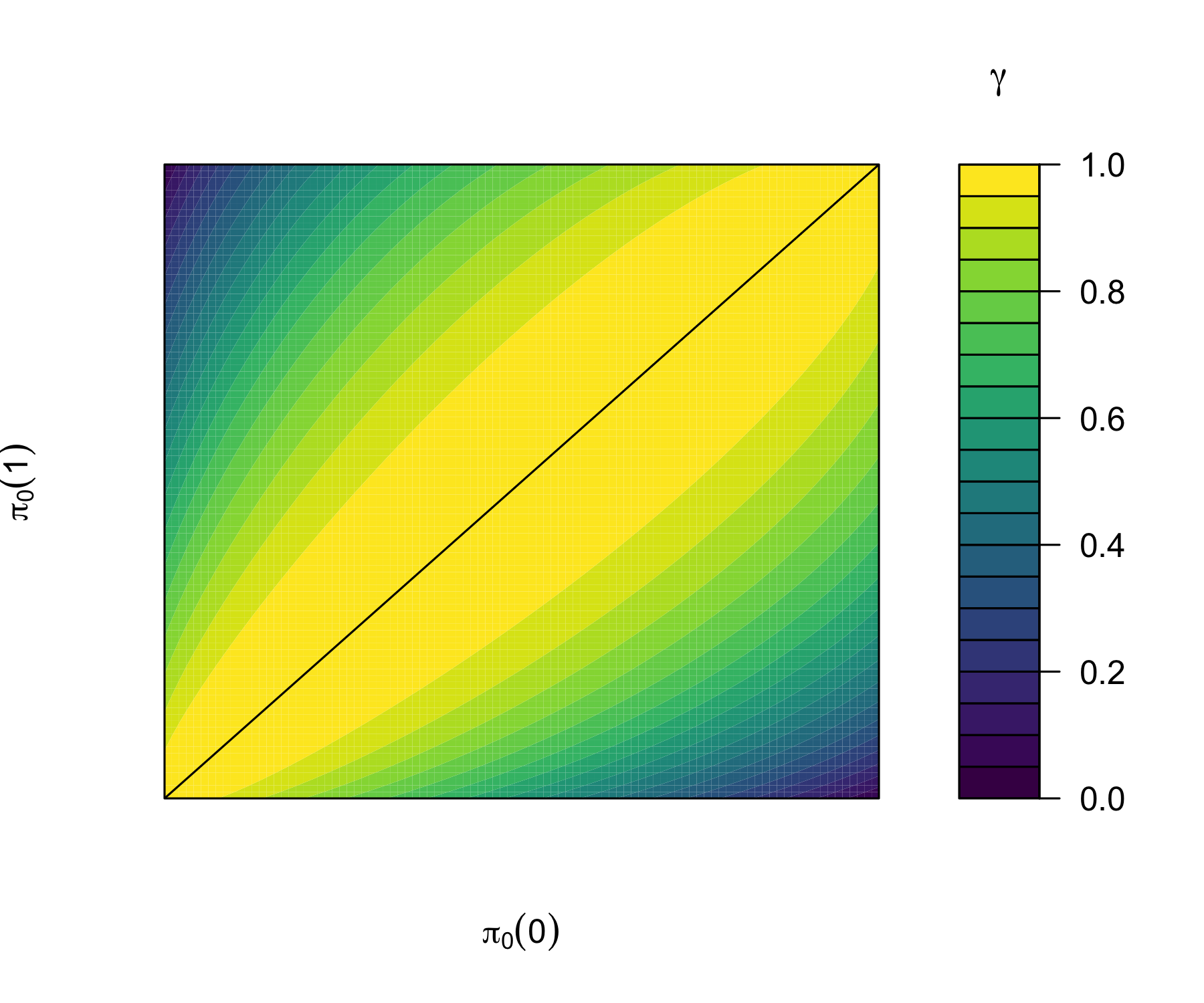}}
	\end{tabular}
	\caption{Illustration of non-collapsibility. For fixed $p(x=1)=0.5$ and $\beta_{t}=1.0$, the marginal log odds-ratio $\gamma$ of treatment becomes closer to $0$ when the difference in the untreated risks $\pi_0(0),\pi_0(1)$ becomes larger.}
	\label{fig:appnc}
\end{figure}

Here we provide an example and intution on what non-collapsibility of the odds-ratio is and why
the difference between the conditional odds-ratio and the marginal odds-ratio
increases when the assocation between $x$ and $y$ becomes greater.
Consider the following data-generating mechanism for binary $X$ with $q(X=1) = 0.5$, binary treatment $T$, and outcome mechanism $\Pr(Y_t=1|x) = \sigma ( b_0(x) + t)$, so that the  \textit{conditional} odds-ratio ($e^1 \approx 2.72$) is constant.
As we will see, depending on how $b_0$ depends on $x$, the \textit{marginal} log odds-ratio $\gamma_{t}$ will vary.
For two settings of $b_0(x)$ we calculate the resulting \textit{marginal} odds-ratio $\gamma_{t}$ in a few simple steps.
The calculations are visualized in Figure \ref{fig:appnc}.
Let $\pi_{t}(x) = \Pr(Y_t=1|x)$: 

\begin{align*}
	\pi_0(0) &= \sigma (b_0 (x=0))  \\
	\pi_0(1) &= \sigma (b_0 (x=1))  \\
	\pi_1(0) &= \sigma (b_0 (x=0) + 1) \\
	\pi_1(1) &= \sigma (b_0 (x=1) + 1) \\
	\pi_0    &= (1-q(x=1)) \pi_0(0) + q(x=1) \pi_0(1) \\
	\pi_1    &= (1-q(x=1)) \pi_1(0) + q(x=1) \pi_1(1) \\
	\eta_0   &= \sigma^{-1} (\pi_0) \\
	\eta_1   &= \sigma^{-1} (\pi_1) \\
	\gamma_{t} &= \eta_1 - \eta_0
\end{align*}

This leads to the following numerical results in Table \ref{tab:appncnum} where we see that $\beta_{t} > \gamma_{t} > 0$ and $\gamma_{t} \to 0$ when the difference between $\pi_0(0),\pi_0(1)$ becomes bigger, despite $\beta_{t}=1$ remaining constant.

\begin{table}[h]
	\centering
	\begin{tabular}{llllllllllll}
		setting & $x$ & $\eta_0(x)$ & $\eta_1(x)$ & $\beta_{t}$ & $\pi_0(x)$ & $\pi_1(x)$ & $\pi_0$ & $\pi_1$ & $\eta_0$ & $\eta_1$ & $\gamma_{t}$ \\
		\hline
		\multirow{2}{*}{a} & \multicolumn{1}{l}{0} &\multicolumn{1}{l}{-1.5} &\multicolumn{1}{l}{-0.5} &\multicolumn{1}{l}{1} &\multicolumn{1}{l}{0.182} &\multicolumn{1}{l}{0.378} &\multirow{2}{*}{0.402} &\multirow{2}{*}{0.598} &\multirow{2}{*}{-0.395} &\multirow{2}{*}{0.395} &\multirow{2}{*}{0.791} \\\cline{2-7}
				   & \multicolumn{1}{l}{1} &\multicolumn{1}{l}{0.5} &\multicolumn{1}{l}{1.5} &\multicolumn{1}{l}{1} &\multicolumn{1}{l}{0.622} &\multicolumn{1}{l}{0.818} & & & & & \\\cline{2-7}
		\hline
		\multirow{2}{*}{b} & \multicolumn{1}{l}{0} &\multicolumn{1}{l}{-3.5} &\multicolumn{1}{l}{-2.5} &\multicolumn{1}{l}{1} &\multicolumn{1}{l}{0.029} &\multicolumn{1}{l}{0.076} &\multirow{2}{*}{0.477} &\multirow{2}{*}{0.523} &\multirow{2}{*}{-0.093} &\multirow{2}{*}{0.093} &\multirow{2}{*}{0.186} \\\cline{2-7}
				   & \multicolumn{1}{l}{1} &\multicolumn{1}{l}{2.5} &\multicolumn{1}{l}{3.5} &\multicolumn{1}{l}{1} &\multicolumn{1}{l}{0.924} &\multicolumn{1}{l}{0.971} & & & & & \\\cline{2-7}
		\hline
	\end{tabular}
	\caption{}
	\label{tab:appncnum}
\end{table}

\FloatBarrier
\FloatBarrier

\subsection{Consistency of marginally constrained models}
\label{app:proofconsistency}

Here we prove Theorem \ref{th:constr}.
First we state the formal version of the theorem: 

\begin{theorem}{(formal version)}
	Given binary treatment $T$, binary outcome $Y$ and covariate $X$.
	Assume strong ignorability $(Y_1,Y_0) \indep T|X$, positivity $0<q(T|X)<1$ and consistency $Y_t=Y$ if $T=t$.
	Given a model $p_{\theta}(y|t,x)$ indexed by parameter $\theta \in \Theta$, and assuming $\exists \theta^* \in \Theta, \Pr(Y_t=1|X=x) = p_{\theta^*}(y|t,x)$.
	Denote the sample log-likelihood $L_n: \Theta \to \mathbb{R}$, marginal log odds-ratio $\gamma = \sigma^{-1}(\Pr(Y_1=1)) - \sigma^{-1}(\Pr(Y_0=1))$, sample marginalizer $M_n: \Theta \to \mathbb{R}$:
	\begin{equation}
		M_n(\theta) = \sigma^{-1}(\frac{1}{n} \sum_{i=1}^n p_{\theta}(Y=1|1,x_i) ) - \sigma^{-1}(\frac{1}{n} \sum_{i=1}^n p_{\theta}(Y=1|0,x_i))
	\end{equation}
	 and $0 < \lambda < \infty$.
	Assume that $L_n$ converges uniformly in probability to $L : \Theta \to \mathbb{R}$ and $M_n$ to $M : \Theta \to \mathbb{R}:$

	\begin{equation}
		M(\theta) = \sigma^{-1}(\EX p_{\theta}(Y=1|1,X)) - \sigma^{-1}(\EX p_{\theta}(Y=1|0,X))
	\end{equation}

	The uniform convergence in probability means:

	\begin{align*}
		\sup_{\theta \in \Theta} |L_n(\theta) - L(\theta)| &\overset{p}{\to} 0 \\
		\sup_{\theta \in \Theta} |M_n(\theta) - M(\theta)| &\overset{p}{\to} 0 
	\end{align*}

	Additionally, assume strong identifiability, namely that for every $\epsilon > 0$ the Kullback-Leibler (KL) divergence between the data distribution and model distribution:

	\begin{equation*}
		\inf_{\theta: | \theta - \theta^* | \geq \epsilon} \text{KL}(p_{\theta^*}(y|t,x),p_{\theta}(y|t,x)) > 0
	\end{equation*}

	Then:
	\begin{equation*}
		\theta_n = \arg \max_{\theta \in \Theta} \left[ L_n(\theta) - \lambda ( M_n(\theta) - \gamma)^2 \right]
	\end{equation*}

	 is a consistent estimator of $\theta^*$ so the model matches $\Pr(Y_t=1|X)$
	
\end{theorem}

To prove this theorem we first prove that $\Pr(Y_t=1|X) = \Pr(Y|T,X)$ (causal identifiability) and then consistency of the constrained estimator for the observational distribution.

\begin{proof}
	A standard causal inference result is that under strong ignorability, positivity and (causal) consistency, $\Pr(Y_t=1|X)=\Pr(Y=1|T,X)$. Restating the proof:

	\begin{align*}
		\Pr(Y_t=1|X) &= \Pr(Y_t=1|X,t) \\
			   &= \Pr(Y=1|X,t) \\
	\end{align*}

	The first equality follows from the strong ingorability assumption $(Y_0,Y_1) \indep T | X$,
	the second follows from the consistency assumption $Y_t = Y, T=t$.
	The right hand side now only contains observable quantities.
	We now prove that the constrained estimator is a consistent estimator of $\Pr(Y_t=1|X)$.

	Writing the population objective and its empirical counterpart:
	\begin{align}
		\mathcal{L}(\theta) &= \EX \log p_{\theta}(y_i|x_i,t_i) - \lambda ( M(\theta) - \gamma)^2 \\
		\mathcal{L}_n(\theta) &= \frac{1}{n}\sum_{i=1}^n \log p_{\theta}(y_i|x_i,t_i) - \lambda ( M_n(\theta) - \gamma)^2 
	\end{align}
	
	We first introduce a new objective:

	\begin{align}
		\label{eq:rn}
		R_n^{\theta^*}(\theta) &= \mathcal{L}_n(\theta^*) - \mathcal{L}_n(\theta) \\
				       &= \frac{1}{n}\sum_{i=1}^n \log \frac{p_{\theta^*}(y_i|x_i,t_i)}{p_{\theta}(y_i|x_i,t_i)} - \lambda ( M_n(\theta^*) - \gamma)^2 + \lambda ( M_n(\theta) - \gamma)^2
	\end{align}

	Because the first term in \ref{eq:rn} is constant it is clear that 

	\begin{equation*}
		\hat{\theta} = \arg \min_{\theta \in \Theta} R_n(\theta) \Longleftrightarrow \hat{\theta} = \arg \max_{\theta \in \Theta} \mathcal{L}_n(\theta)
	\end{equation*}

	The population version of $R_n^{\theta^*}$ is:
	\begin{align}
		\label{eq:re}
		R^{\theta^*}(\theta) &= \EX_{t,x} \EX_{y|t,x} \log \frac{p_{\theta^*}(y|x,t)}{p_{\theta}(y|x,t)} - \lambda ( M(\theta^*) - \gamma)^2 + \lambda ( M(\theta) - \gamma)^2 \\
				     &\overset{(i)}= \EX_{t,x} \text{KL}(p_{\theta^*}(y|x,t) || p_{\theta}(y|x,t)) - \lambda ( M(\theta^*) - \gamma)^2 + \lambda ( M(\theta) - \gamma)^2 \\
				     &\overset{(ii)}= \EX_{t,x} \text{KL}(p_{\theta^*}(y|x,t) || p_{\theta}(y|x,t)) + \lambda ( M(\theta) - \gamma)^2
	\end{align}
	Where $(i)$ is substituting the definition of the KL-divergence and equality $(ii)$ holds as $\gamma = M(\theta^*)$ by definition.

	Lemma \ref{lem:sum} gives that uniform convergence in probability of $L_n$ and $M_n$ imply uniform convergence in probability of $\mathcal{L}_n$ and thus $R_n$ as $R_n$ is just a constant minus $\mathcal{L}_n$.
	We now use this convergence to prove that $R_n^{\theta^*} \overset{p}\to 0$ as $n \to \infty$, which will prove the consistency of the constrained estimator.
	Denote $\hat{\theta}_n$ as the maximizer of $\mathcal{L}_n$, because of this, it must be that

	\begin{equation*}
		R_n^{\theta^*}(\hat{\theta}_n) = \mathcal{L}_n(\theta^*) - \mathcal{L}_n(\hat{\theta}_n) \leq 0
	\end{equation*}

	We can now bound $R^{\theta^*}(\hat{\theta}_n)$ and show that it converges to $0$ in probability:

	\begin{align}
		R^{\theta^*}(\hat{\theta}_n) = R^{\theta^*}(\hat{\theta}_n) - R_n^{\theta^*}(\hat{\theta}_n) + R_n^{\theta^*}(\hat{\theta}_n) \overset{(i)}\leq R^{\theta^*}(\hat{\theta}_n) - R_n^{\theta^*}(\hat{\theta}_n) \overset{p}\to 0
	\end{align}
	The inequality $(i)$ holds as $R_n^{\theta}(\hat{\theta}) \leq 0$ and the convergence is given by uniform convergence of $R_n$.
	To show that the KL-divergence goes to zero, note that:
	\begin{align}
		R^{\theta^*}(\hat{\theta}_n) &= \underbrace{\EX_{t,x} \text{KL}(p_{\theta^*}(y|x,t) || p_{\hat{\theta}_n}(y|x,t))}_{A_n} + \underbrace{\lambda ( M(\hat{\theta}_n) - \gamma)^2}_{B_n} \overset{p}\to 0 
	\end{align}
	$A_n$ is non-negative because the KL-divergence is non-negative and $B_n$ is non-negative as well.
	For all $n$, $A_n$ is lower bounded by $0$ and upper bounded by $R_n$,
	by application of the squeeze theorem adapted for convergence in probability,
	$R_n \overset{p}\to 0 \implies A_n \overset{p} \to 0$.
	For a short proof on how the squeeze theorem for sequences translates to convergence in probability, see Lemma \ref{lem:squeeze} below.
	The assumption of strong identifiability gaurantees that $\theta \to \theta^*$ as the KL-divergence goes to zero, proving the consistency of the constrained estimator.

\end{proof}

\begin{lemma}
	\label{lem:simple-sum}
	Let $A_n(\theta),B_n(\theta) \in \mathbb{R}$ be sequences of real-valued random variables for parameter $\theta$ and $A(\theta),B(\theta)$ be random variables in $\mathbb{R}$. If both
	\begin{align}
		\sup_{\theta} | A_n(\theta) - A(\theta) | &\overset{p}\to 0  \nonumber \\
		\sup_{\theta} | B_n(\theta) - B(\theta) | &\overset{p}\to 0
		\label{eq:simple-sum-assumption}
	\end{align}
	then,
	\begin{align*}
	\sup_{\theta} | A_n(\theta) + B_n(\theta) - A(\theta) - B(\theta)|  &\overset{p}\to 0.
	\end{align*}
\end{lemma}

\begin{proof}
	Define $C_n(\theta) = A_n(\theta) - A(\theta)$ and $D_n(\theta) = B_n(\theta) - B(\theta)$. By the definition of convergence in probability, the goal is to show that 
	\begin{align*}
		\lim_{n \to \infty} \Pr(\sup_{\theta} | C_n(\theta) +  D_n(\theta| > \epsilon) = 0
	\end{align*}
	Now construct an upper bound on the probability 
	\begin{align}
		&\Pr(\sup_{\theta} | C_n(\theta) + D_n(\theta) | > \epsilon) 
		\nonumber \\
		&\leq \Pr(\sup_{\theta} | C_n(\theta)| +  |D_n(\theta)| > \epsilon) 
		\nonumber \\
		&\leq \Pr(\sup_{\theta_C} | C_n(\theta_C)| +  \sup_{\theta_D} |D_n(\theta_D)| > \epsilon) 
		\nonumber \\
		&\leq \Pr(\sup_{\theta_C} | C_n(\theta_C)| > \nicefrac{\epsilon}{2} \quad \text{or} \quad \sup_{\theta_D} |D_n(\theta_D)| > \nicefrac{\epsilon}{2}) 
		\\
		&\leq \Pr(\sup_{\theta} | C_n(\theta)|  >  \nicefrac{\epsilon}{2}) + \Pr(\sup_{\theta} |D_n(\theta)| > \nicefrac{\epsilon}{2})
		\nonumber \\
		&= \Pr(\sup_{\theta} |A_n(\theta) - A(\theta)| > \nicefrac{\epsilon}{2}) 
		+ \Pr(\sup_{\theta} | B_n(\theta) - B(\theta) |  >  \nicefrac{\epsilon}{2}).
		\label{eq:simple-sum-upper-bound}
	\end{align}
	By the assumption of convergence in probability in Equation \ref{eq:simple-sum-assumption}, 
	\begin{align*}
		\lim_{n \to \infty} \Pr(\sup_{\theta} |A_n(\theta) - A(\theta)| > \nicefrac{\epsilon}{2}) 
		+ \Pr(\sup_{\theta} | B_n(\theta) - B(\theta) |  >  \nicefrac{\epsilon}{2}) = 0.
	\end{align*}
	This limit shows that $\Pr(\sup_{\theta} | C_n(\theta) + D_n(\theta) | > \epsilon)$ goes to zero via the squeeze theorem because the upper bound, Equation \ref{eq:simple-sum-upper-bound}, goes to zero and the lower bound on probabilities of zero, thus showing the desired convergence in probability.
\end{proof}

\begin{lemma}
	\label{lem:square}
	Let $M_n(\theta) \in \mathbb{R}$ be a sequence of real-valued random variables for parameter $\theta \in \Theta$ and $M(\theta),\gamma \in \mathbb{R}$ for $\theta \in \Theta$.
	If $\sup_{\theta \in \Theta}|M_n(\theta) - M(\theta)| \overset{p}\to 0$ then $\sup_{\theta \in \Theta}|(M_n(\theta) - \gamma)^2 - (M(\theta) - \gamma)^2| \overset{p}\to 0$
\end{lemma}

\begin{proof}

	Define $\xi_n (\theta)= M_n (\theta)- M(\theta)$, then:

	\begin{equation}
		\label{eq:xi}
		\sup_{\theta \in \Theta}|M_n(\theta) - M(\theta)| \overset{p}\to 0 \implies \sup_{\theta \in \Theta}|\xi_n(\theta)| \overset{p}\to 0
	\end{equation}

	We have that

	\begin{align}
		(M_n - \gamma)^2 - (M - \gamma)^2 =& (M + \xi_n - \gamma)^2 - (M - \gamma)^2 \\
		=& M^2 + \xi_n^2 + \gamma^2 + 2 M \xi_n - 2 \gamma \xi_n - 2 M \gamma \\
		 &- M^2 - \gamma^2 + 2 M \gamma \\
		=& \xi_n^2 + 2 \xi_n(M - \gamma)
	\end{align}

	We need to show that 

	\begin{equation*}
		\Pr(\sup_{\theta \in \Theta}| \xi_n(\theta)^2 + 2\xi_n(\theta) (M(\theta) - \gamma) | > \epsilon) \underset{n\to \infty}{=} 0
	\end{equation*}

	This expression is the sum of two sequences with $\xi_n$.
	By Lemma~\ref{lem:simple-sum}, our desired result follows if both:

	\begin{align*}
		\sup_{\theta \in \Theta} |\xi_n(\theta)^2| &\overset{p}\to 0\\
		\sup_{\theta \in \Theta} |2 \xi_n(\theta)(M(\theta) - \gamma)| &\overset{p}\to 0
	\end{align*}

	We can bound the term with $\xi_n$ using \ref{eq:xi} and the fact that $M(\theta)-\gamma \in \mathbb{R}, \forall \theta \in \Theta$.
	For $|\xi_n|^2$ to converge to $0$, note that:

	\begin{align*}
		\Pr(\sup_{\theta \in \Theta}|\xi_n(\theta)|^2 > \epsilon) = \Pr(\sup_{\theta \in \Theta}[|\xi_n(\theta)||\xi_n(\theta)|] > \epsilon) = \Pr(\sup_{\theta \in \Theta}|\xi_n(\theta)| > \sqrt{\epsilon})
	\end{align*}

	Again, applying \ref{eq:xi} gives the required result.
\end{proof}

\begin{lemma}
	\label{lem:sum}
	Let $L_n(\theta),M_n(\theta) \in \mathbb{R}$ be sequences of real-valued random variables for $\theta \in \Theta$ and $L(\theta),M(\theta),\gamma \in \mathbb{R}$, and let $0< \lambda < \infty$.
	Define
	\begin{align*}
		\mathcal{L}_n &= L_n - \lambda (M_n - \gamma)^2 \\
		\mathcal{L}   &= L - \lambda (M - \gamma)^2 \\
	\end{align*}
	Then if $L_n \overset{p}\to L$ uniformly and $M_n \overset{p}\to M$ uniformly, then $\mathcal{L}_n \overset{p}\to \mathcal{L}$ uniformly
\end{lemma}

\begin{proof}
	By Lemma \ref{lem:square} we have that
	\begin{equation*}
		\sup_{\theta} |M_n(\theta) - M(\theta)| \overset{p}\to 0 \implies \sup_{\theta \in \Theta}|(M_n(\theta) - \gamma)^2 - (M(\theta) - \gamma)^2 | \overset{p}\to 0
	\end{equation*}
	By uniform convergence in probability we have that also
	\begin{align*}
		\sup_{\theta} | L_n(\theta) - L(\theta) | &\overset{p}\to 0
	\end{align*}

	This gives us
	\begin{align*}
		\sup_{\theta} |\mathcal{L}_n(\theta) - \mathcal{L}(\theta)| &= \sup_{\theta} |L_n(\theta) - \lambda (M_n(\theta) - \gamma)^2 - L + \lambda (M(\theta) - \gamma)^2 | \\
									    &\overset{i}\leq \sup_{\theta} |L_n(\theta) - L(\theta)|  +  \lambda \sup_{\theta} | (M(\theta) - \gamma)^2 - (M_n(\theta) - \gamma)^2 | \overset{p}{\to} 0
	\end{align*}
	$(i)$ follows from the triangle inequality and the final convergence follows from $0 < \lambda < \infty$ and the uniform convergence of the individual terms.
\end{proof}

\begin{lemma}
	\label{lem:squeeze}
	If $X_n \leq Y_n \leq Z_n$ for every $n$ and $X_n \overset{p}\to L$ and $Z_n \overset{p}\to L$, then $Y_n \overset{p}\to L$
\end{lemma}

\begin{proof}
	We want to show that for any $\epsilon > 0$, we have 
	\begin{equation*}
		\lim_{n \to \infty} \Pr(|Y_n - L| > \epsilon) = 0
	\end{equation*}
	Note that
	\begin{equation*}
		X_n \leq Y_n \leq Z_n
	\end{equation*}
	implies
	\begin{equation*}
		|Y_n - L| \leq |X_n - L| + |Z_n - L|
	\end{equation*}
	so 
	\begin{equation*}
		\Pr(|Y_n - L|>\epsilon) \leq \Pr(|X_n - L| + |Z_n - L| > \epsilon)
	\end{equation*}
	The right-hand side can be made arbitrarily small by convergence of $X_n$ and $Z_n$, proving what we want to show.
\end{proof}

\subsection{The offset model is not a consistent estimator in the presence of unobserved confounding}
\label{app:proofnotstationary}

We now prove that optimization over the offset model family in Equation \ref{eq:offset_family} and a known conditional odds-ratio $\beta_t$ leads to an asymptotically biased estimator of $\Pr(Y_t|x)$ in the presence of unobserved confounding in a simple example introduced in the main text (\ref{sec:example1}).
In this example, $\Pr(Y_t|x) = \Pr(Y_t)$, or equivalently, there is no covariate $X$.
This also implies that here CATE=ATE, so we will use ATE instead.
Given the offset model family from Equation \ref{eq:offset_family}, a natural parameterization of $g(t,x)=g(t)$ in the context of this example is $g(t;b_0) = \sigma (b_0 + \beta_t t), b_0 \in \mathbb{R}$.
Again, we are assuming that $\beta_t$ is given a-priori and is not estimated.
We first derive an expression for the expected log-likelihood as a function of $b_0$
under the observational distribution in this example.
Then we show that the ground truth solution $\beta_0$ is not a stationary point, proving our claim.
Writing

\begin{align*}
	p_u = &p(u=1) \\
	p_{{t}'u'} = &p({t}={t}',u=u') = p({t}={t}'|u=u')p(u=u')\\
	\pi_{{t}'u'} = &p(y=1|{t}={t}',u=u')
\end{align*}

Then the data generating mechanism is: 

\begin{equation*}
	t,u \sim \mathcal{B}(p_{{t}'u'}), y \sim \mathcal{B}(\pi_{{t}u})
\end{equation*}

The ground truth solutions $\beta_0$ and $\beta_{t}$ are:

\begin{align}\label{eq:gt}
	\Pr(Y_0=1) &= (1 - p_u) \pi_{00} + p_u \pi_{01} = \sigma (\beta_0) \\
	\Pr(Y_1=1) &= (1 - p_u) \pi_{10} + p_u \pi_{11} = \sigma (\beta_0 + \beta_{t})
\end{align}

The Bernoulli log-likelihood is 

\begin{equation*}
	l(y|{t},b_0) = y \log \sigma(b_0 + \beta_{t} {t}) + (1 - y) \log (1 - \sigma (b_0 + \beta_{t} t))
\end{equation*}

In offset models $\beta_{t}$ is assumed given a priori and $b_0$ is the only parameter,
resulting in the following expression for $L(b_0)$:

\begin{align*}
	L(b_0) = & p_{00} \left[ \pi_{00} \log \sigma (b_0) + (1 - \pi_{00}) \log(1 - \sigma ( b_0)) \right] \\
	           + & p_{01} \left[ \pi_{01} \log \sigma (b_0) + (1 - \pi_{01}) \log(1 - \sigma ( b_0)) \right] \\ 
	           + & p_{10} \left[ \pi_{10} \log \sigma (b_0 + \beta_{t}) + (1 - \pi_{10}) \log(1 - \sigma ( b_0 + \beta_{t})) \right] \\ 
	           + & p_{11} \left[ \pi_{11} \log \sigma (b_0 + \beta_{t}) + (1 - \pi_{11}) \log(1 - \sigma ( b_0 + \beta_{t})) \right]
\end{align*}

Taking the derivative with respect to $b_0$, noting that $(\log \sigma(x))' = 1 - \sigma (x)$, we get:

\begin{align}
	\frac{\partial L}{\partial b_0} = & p_{00} \left[ \pi_{00} (1 - \sigma (b_0)) - (1 - \pi_{00}) \sigma ( b_0) \right] \label{eq:dl}\\
	+ & p_{01} \left[ \pi_{01} (1 - \sigma (b_0)) - (1 - \pi_{01}) \sigma ( b_0) \right] \nonumber \\ 
	+ & p_{10} \left[ \pi_{10} (1 - \sigma (b_0 + \beta_{t})) - (1 - \pi_{10}) \sigma ( b_0 + \beta_{t}) \right] \nonumber \\ 
	+ & p_{11} \left[ \pi_{11} (1 - \sigma (b_0 + \beta_{t})) - (1 - \pi_{11}) \sigma ( b_0 + \beta_{t}) \right] \nonumber 
\end{align}

We now plug in the ground truth solutions for $\beta_0,\beta_{t}$ (Equations \ref{eq:gt}).

\begin{align*}
	\frac{\partial L}{\partial b_0}(b_0 = \beta_0) = & p_{00} \left[ \pi_{00} (1 - p_u \pi_{01} - (1 - p_u) \pi_{00}) - (1 - \pi_{00}) (p_u \pi_{01} + (1-p_u) \pi_{00} ) \right] \\
	+ & p_{01} \left[ \pi_{01} (1 - p_u \pi_{01} - (1 - p_u) \pi_{00}) - (1 - \pi_{01}) (p_u \pi_{01} + (1-p_u) \pi_{00} ) \right] \\ 
	+ & p_{10} \left[ \pi_{10} (1 - p_u \pi_{11} - (1 - p_u) \pi_{10}) - (1 - \pi_{10}) (p_u \pi_{11} + (1-p_u) \pi_{10} ) \right] \\ 
	+ & p_{11} \left[ \pi_{11} (1 - p_u \pi_{11} - (1 - p_u) \pi_{10}) - (1 - \pi_{11}) (p_u \pi_{11} + (1-p_u) \pi_{10} ) \right]
\end{align*}

Removing terms that cancel out in each line results in 

\begin{align*}
        \frac{\partial L}{\partial b_0}(b_0 = \beta_0) = & p_{00} \left[ p_u (\pi_{00} - \pi_{01}) \right] \\
	+ & p_{01} \left[ (1 - p_u) (\pi_{01} - \pi_{00}) \right] \\ 
	+ & p_{10} \left[ p_u (\pi_{10} - \pi_{11}) \right] \\ 
	+ & p_{11} \left[ (1 - p_u) (\pi_{11} - \pi_{10}) \right]
\end{align*}

Substituting back $p_{{t}'u'} = p({t}={t}'|u=u')p(u=u')$:

\begin{align*}
        \frac{\partial L}{\partial b_0}(b_0 = \beta_0) = & p({t}=0|u=0)(1-p_u) \left[ p_u (\pi_{00} - \pi_{01}) \right ] \\
	+ & p({t}=0|u=1)p_u \left[ (1-p_u) (\pi_{01} - \pi_{00}) \right ] \\ 
	+ & p({t}=1|u=0)(1-p_u) \left[ p_u (\pi_{10} - \pi_{11}) \right ] \\ 
	+ & p({t}=1|u=1)p_u \left[ (1-p_u) (\pi_{11} - \pi_{10}) \right ]
\end{align*}

Factoring out $p_u(1-p_u)$ and re-arranging we arrive at our result:

\begin{align*}
	\frac{\partial L}{\partial b_0}(b_0 = \beta_0) = p_u (1-p_u) \bigl[ & (\pi_{01} - \pi_{00}) \left( p({t}=0|u=1) - p({t}=0|u=0) \right) + \\
									    & (\pi_{11} - \pi_{10}) \left( p({t}=1|u=1) - p({t}=1|u=0) \right) \bigr]
\end{align*}

If there is no confounding ($\pi_{t0} = \pi_{t1}$ or $p(t|u=0) = p(t|u=1)$) this expression is zero,
but in general it is not which means that the ground truth solution $\beta_0$ is not an optimum of the expected log-likelihood in the observational data distribution.
This proves our claim that the offset model does not recover $\Pr(Y_t)$ in the presence of confounding. $\square$

Of note, the fact that the offset model does not estimate $\Pr(Y_t)$ does not automatically imply that the \gls{ate} is not correctly estimated as there may be another $b_0' \neq \beta_0$ such that $\widehat{\text{ATE}}(b_0 = b_0') = {\text{ATE}}(b_0 = \beta_0)$.
To investigate this, assume that for some $\beta_0 = a$ and $\beta_{t} = c$  we have that:

\begin{align*}
	\delta := & \sigma(a + c) - \sigma(a) \\
	  = & \frac{e^{a + c}}{1 + e^{a+c}} - \frac{e^a}{1 + e^{a}}
\end{align*}

Again, treating $\beta_{t}$ as fixed, we will now prove that this equation has at most two solutions for $b_0=a$ by noting that:

\begin{align*}
 \frac{e^{a + c}}{1 + e^{a+c}} - \frac{e^a}{1 + e^{a}} &= \frac{e^{a + c}(1+e^a) - (1+e^{a+c})e^a}{(1 + e^{a+c})(1+e^a)} \\
	&= \frac{e^{a}(e^c-1)}{(1 + e^{a+c})(1+e^a)}
\end{align*}

Introducing $y:=e^a$ and cross-multiplying we get:

\begin{align*}
	\delta &= \frac{y(e^c-1)}{(1 + e^c y)(1+y)} \iff \\
	\delta (1 + e^c y)(1 + y) &= y (e^c-1) = \\
	\delta + \delta (1 + e^c) y + \delta e^c y^2 &= y (e^c-1) \iff \\
	\delta e^c y^2 + \left( \delta (1 + e^c ) - e^c + 1 \right) y + \delta &= 0
\end{align*}

Depending on the values of $\delta$ and $c$ this quadratic equation in $y$ has 0, 1 or 2 real-valued solutions, yielding 0, 1 or 2 real-valued solutions for $a = \log y = b_0$.
This implies that there exists utmost one alternative solution $b_0' \neq \beta_0$ such that $\widehat{\text{ATE}}(b_0 = b_0') = {\text{ATE}}(b_0 = \beta_0)$.

In fact, we can explicitly compute this alternative solution by exploiting the symmetry of the sigmoid function: $\sigma(x) = 1 - \sigma(-x)$.
Whenever it is true that:

\begin{equation*}
	\sigma(\beta_0 + \beta_{t}) - \sigma(\beta_0) = \delta
\end{equation*}

It must simultaneously be true that, writing $\beta'_0 := -(\beta_0 + \beta_{t})$:

\begin{align*}
	\sigma(\beta'_0 + \beta_{t}) - \sigma(\beta'_0) &= \\
	\sigma(-(\beta_0 + \beta_{t}) + \beta_{t}) - \sigma(-(\beta_0 + \beta_{t})) &= \\
	\sigma(- \beta_0) - \sigma(-(\beta_0 + \beta_{t})) &= \\
	(1 - \sigma(\beta_0)) - (1 - \sigma(\beta_0 + \beta_{t})) &= \\
	\sigma(\beta_0 + \beta_{t}) - \sigma(\beta_0) &= \delta
\end{align*}

This means that except in the trivial case when $\beta_0 = \beta_{t} = 0$ there \textit{always} exists a second solution $\beta'_0$ that has the same \gls{ate} $\delta$ but a different $\Pr(Y_t)$.
We can check whether this coincidentally coincides with the maximum likelihood solution for $b_0$ in the offset model on the observational data by plugging in $\beta'_0 := -(\beta_0 + \beta_{t})$ in the expression of the gradient of the likelihood (Equation \ref{eq:dl}).
Again we remove terms that cancel out and substitute back $p_{{t}'u'} = p({t}={t}'|u=u')p(u=u')$ to arrive at:

\begin{align}
	\frac{\partial L}{\partial b_0}(b_0 = \beta'_0) & = p_u (1-p_u) \bigl( p({t}=0|u=1) - p({t}=0|u=0) \bigr) \bigl( (\pi_{10} - \pi_{11}) + (\pi_{01} - \pi_{00}) \bigr) \\
	& + p_u \bigl( (\pi_{11} - \pi_{10}) + (\pi_{01} - \pi_{00}) \bigr) \\
	& + 2 \pi_{10} + \pi_{11} - 1
\end{align}

Analyzing this expression line-by-line we see that the first two lines are non-zero in general when there is confounding such that $p({t}=0|u=1) \neq p({t}=0|u=0)$ and $\pi_{{t}1} \neq \pi_{{t}0}$.
The last line is also non-zero in general as $\pi_{{t}u}$ are free parameters.
This shows that the offset model is also an asymptotically biased estimator of the \gls{ate}.

\subsection{Analytical solution of offset model with binary covariate}
\label{app:analytical_solution}

Assume a binary treatment $t$, covariate $x$ and outcome $y$.
Write:

\begin{align}
	\label{eq:betat}
	\beta_t^0 &= \sigma^{-1}(\Pr(Y_1=1|x=0)) - \sigma^{-1}(\Pr(Y_0=1|x=0)) \\
	\beta_t^1 &= \sigma^{-1}(\Pr(Y_1=1|x=1)) - \sigma^{-1}(\Pr(Y_0=1|x=1))
\end{align}

The offset assumption states that $\beta_t^0 = \beta_t^1 = \beta_t$.

\subsubsection{Offset solution}

Maximizing the likelihood over a model class is equivalent to minimizing the Kullback-Leibler divergence between the actual distribution and the model distribution.
We find an analytical solution for the offset models by writing the estimation objective as 
minimizing the Kullback-Leibler divergence between the model distribution and the observed data distribution with the additional constraints above \ref{eq:betat}.
Writing:

\begin{itemize}
	\item $p_{tx}$ for estimated probabilities of $\Pr(Y=1|t,x)$, such that $p(y|t,x) = (1-y)(1-p_{tx}) + y p_{tx}$
	\item $q_{tx}$ for actual (observational) probabilities of $\Pr(Y=1|t,x)$, such that $q(y|t,x) = (1-y)(1-q_{tx}) + y q_{tx}$
	\item $q(t,x)= \Pr(T=t,X=x)$, the observational joint probability of $T,X$
\end{itemize}

The criterion is:

\begin{align}
	\label{eq:kl}
	L &= \EX_{t,x} \EX_{y|t,x} \log \frac{q(y|t,x)}{p(y|t,x)} + \lambda_0 \left( \sigma^{-1}(p_{10}) - \sigma^{-1}(p_{00}) - \beta_t \right) + \lambda_1 \left( \sigma^{-1}(p_{11}) - \sigma^{-1}(p_{01}) - \beta_t \right)
\end{align}

The partial derivatives for all parameters $(p_{00},p_{01},p_{10},p_{11},\lambda_0,\lambda_1)$ are:

\begin{align}
	\label{eq:dlp00}
	\frac{\partial L}{\partial p_{00}} &= \EX_{t,x} \EX_{y|t,x} \frac{\partial}{\partial p_{00}} \left[ \left( \log q(y|t,x) - \log p(y|t,x) \right) + \lambda_0 \left( \sigma^{-1}(p_{10}) - \sigma^{-1}(p_{00}) - \beta_t \right) + \lambda_1 \left( \sigma^{-1}(p_{11}) - \sigma^{-1}(p_{01}) - \beta_t \right) \right] \\
					   &= \EX_{t,x} \EX_{y|t,x} \frac{\partial}{\partial p_{00}} \left[ \left( \log q(y|t,x) - \log (y p_{tx} + (1-y)(1-p_{tx}) \right) \right] - \lambda_0 \frac{\partial}{\partial p_{00}} \sigma^{-1}(p_{00})  \\
					   &= q(0,0) \left[ (1 - q_{00}) \frac{1}{1-p_{00}} - q_{00} \frac{1}{p_{00}} \right]  - \lambda_0 \left( \frac{1}{p_{00}} + \frac{1}{1-p_{00}} \right) \\
					   &= q(0,0) \left[ \frac{(1 - q_{00})p_{00} - q_{00}(1-p_{00})}{p_{00}(1-p_{00})} \right]  - \lambda_0 \left( \frac{(1-p_{00}) + p_{00}}{p_{00}(1-p_{00})} \right) \\
					   &= q(0,0) \left[ \frac{p_{00} - q_{00}}{p_{00}(1-p_{00})} \right]  - \lambda_0 \left( \frac{1}{p_{00}(1-p_{00})} \right) \\
					   &= \frac{q(0,0) \left( p_{00} - q_{00} \right) - \lambda_0}{p_{00}(1-p_{00})} \\ 
	\frac{\partial L}{\partial p_{01}}&= \frac{q(0,1) \left( p_{01} - q_{01} \right) - \lambda_1}{p_{01}(1-p_{01})} \\
	\frac{\partial L}{\partial p_{10}}&= \frac{q(1,0) \left( p_{10} - q_{10} \right) + \lambda_0}{p_{10}(1-p_{10})} \\
	\frac{\partial L}{\partial p_{11}}&= \frac{q(1,1) \left( p_{11} - q_{11} \right) + \lambda_1}{p_{11}(1-p_{11})} \\ 
	\frac{\partial L}{\partial \lambda_0 }&= \sigma^{-1}(p_{10}) - \sigma^{-1}(p_{00}) - \beta_t \\
	\frac{\partial L}{\partial \lambda_1 }&= \sigma^{-1}(p_{11}) - \sigma^{-1}(p_{01}) - \beta_t \\
\end{align}

Setting the gradient to zero to find an extremum of \ref{eq:kl} and requiring $0<p_{tx}<1$ we get:

$0=$
\begin{enumerate}[(a)]
	\item $q(0,0) \left( p_{00} - q_{00} \right) - \lambda_0$
	\item $q(0,1) \left( p_{01} - q_{01} \right) - \lambda_1$
	\item $q(1,0) \left( p_{10} - q_{10} \right) + \lambda_0$
	\item $q(1,1) \left( p_{11} - q_{11} \right) + \lambda_1$
	\item $\sigma^{-1}(p_{10}) - \sigma^{-1}(p_{00}) - \beta_t$
	\item $\sigma^{-1}(p_{11}) - \sigma^{-1}(p_{01}) - \beta_t$
\end{enumerate}

Combining (a) and (c):

\begin{align}
	a + c &= q(0,0) (p_{00} - q_{00} ) + q(1,0) (p_{10} - q_{10}) - \lambda_0 + \lambda_0 = 0 \\
	    0 &= q(0,0) (p_{00} - q_{00} ) + q(1,0) (p_{10} - q_{10}) 
\end{align}

Inserting (e): $p_{10} = \sigma \left( \sigma^{-1}(p_{00}) + \beta_t \right)$ we get

\begin{align}
	     0 &= q(0,0) (p_{00} - q_{00} ) + q(1,0) (\sigma \left( \sigma^{-1}(p_{00}) + \beta_t \right) - q_{10}) \\
\end{align}

Using $\sigma \left( \sigma^{-1}(x) + a \right) = \frac{e^a x}{(e^a-1)x+ 1}$ we get:

\begin{align}
	0 &= q(0,0) (p_{00} - q_{00} ) + q(1,0) \left(\frac{e^{\beta_t} p_{00}}{(e^{\beta_t}-1)p_{00} + 1)} - q_{10} \right) \\
\end{align}

By multiplying both sides with $(e^{\beta_t} - 1)p_{00}+1$ we end up with a quadratic equation in $p_{00}$.
Using some intermediate variables to shorten the notation and applying the same steps to get $p_{01}$, the solution is:

\begin{align*}
	\label{eq:offsetsolution}
	\tau &= e^{\beta_t} \\
	 r_x &= q(1,x) / q(0,x) \\
	 c_x &= -(q(y|0,x) + r_x q(y|1,x)) \\
	 b_x &= 1 + r_x \tau + c_x (\tau - 1) \\
	 p_{0x} &= \frac{-b_x \pm \sqrt{b_x^2+4(\tau - 1) c_x}}{2(\tau -1)}
\end{align*}

The $\pm$ solution is taken to equal sign$(\beta_t)$.
Without loss of generality we assume $\beta_t >0$.
The corresponding solutions $p_{1x}$ are given by $p_{1x} = \sigma(\sigma^{-1}(p_{0x}) + \beta_t)$.

\begin{figure}[htpb]
	\centering
	\includegraphics[width=0.8\paperwidth]{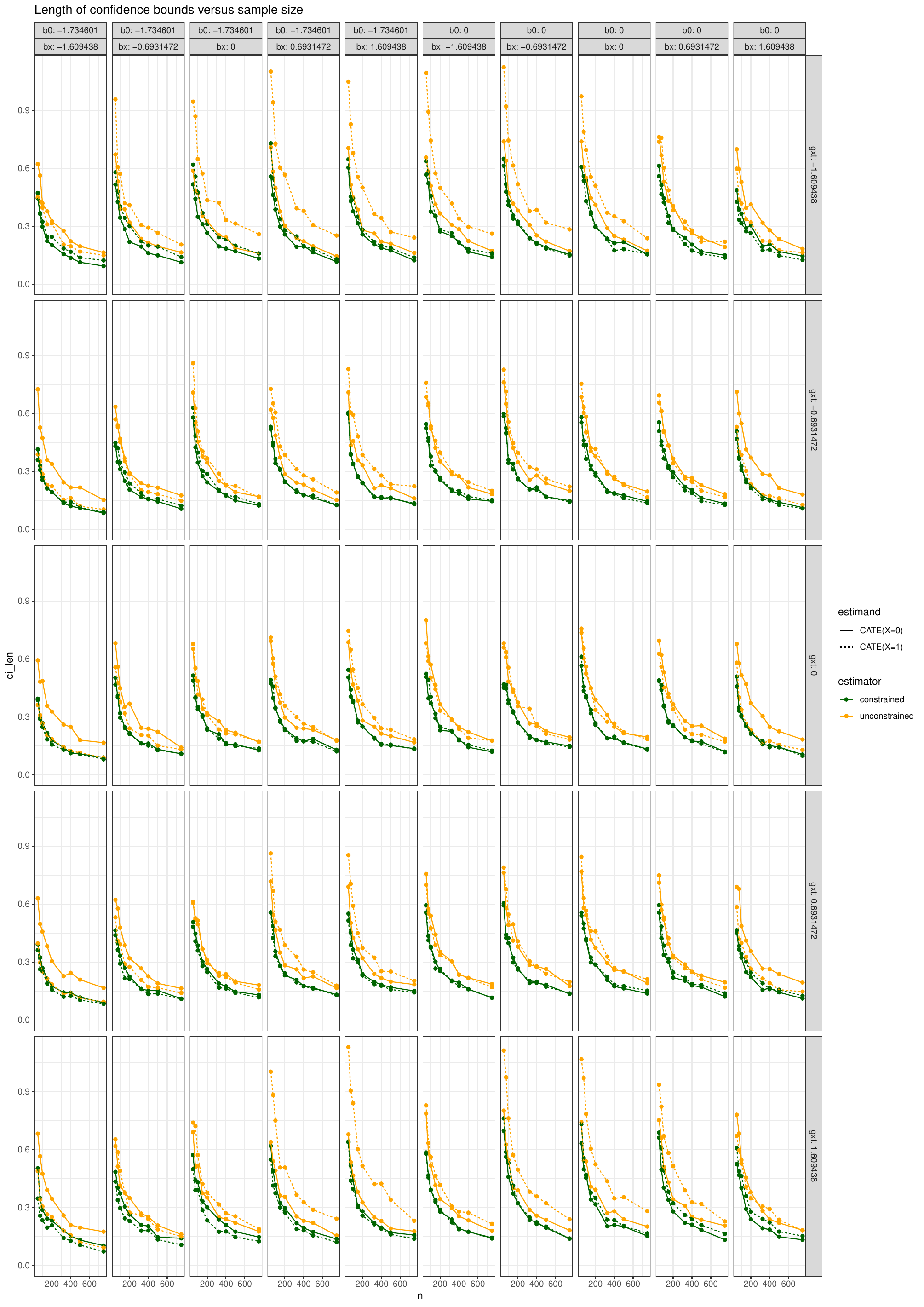}
	\caption{}
	\label{fig:app_ci_vs_n}
\end{figure}

\begin{figure}[htpb]
	\centering
	\includegraphics[width=0.8\paperwidth]{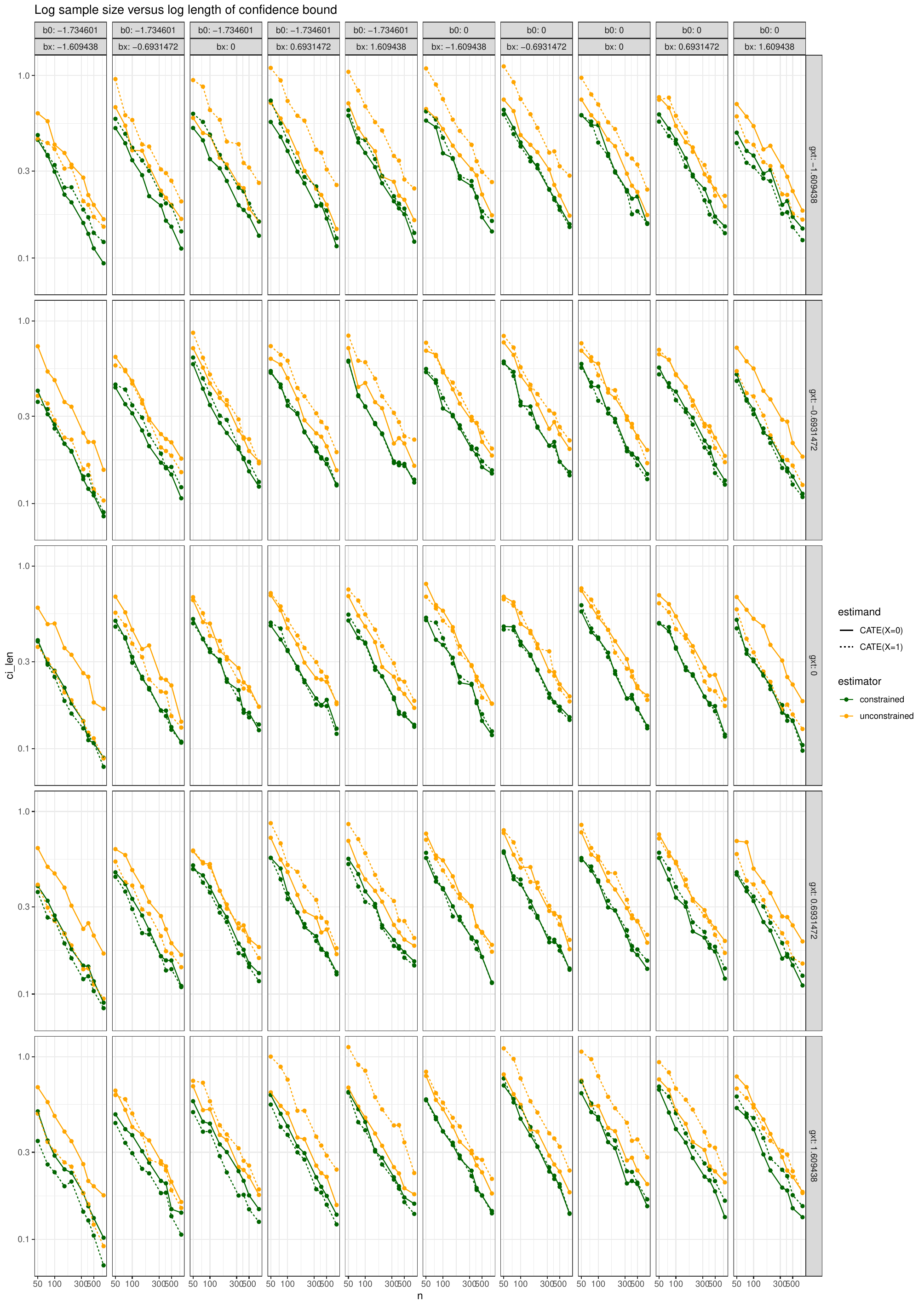}
	\caption{}
	\label{fig:app_log_n_vs_log_ci}
\end{figure}

\pagebreak

\end{document}